\documentclass[useAMS,usenatbib]{mn2e}
\usepackage[dvips]{graphicx}
\usepackage{natbib}
\usepackage{hyperref}
\usepackage{deluxetable}
\usepackage{amssymb}
\usepackage{verbatim}
\title[PS of Isolated NSs II: from radio-pulsars to magnetars]
{Population Synthesis of Isolated Neutron Stars with magneto-rotational evolution II: from radio-pulsars to magnetars}

\author[M.~Gull\'on et al.]
{M. Gull\'on$^1$, J.A.~Pons$^1$,  J. A. Miralles$^1$, D. Vigan\`o$^2$,  N. Rea$^{3,2}$,  R. Perna$^4$\\
$^1$ Departament de F\'isica Aplicada, Universitat d'Alacant, Ap. Correus 99, 03080 Alacant, Spain\\ 
$^2$ Institute of Space Sciences (CSIC-IEEC), Campus UAB, Carrer de Can
Magrans s/n, 08193, Barcelona, Spain \\
$^3$ Anton Pannekoek Institute for Astronomy, University of Amsterdam, Postbus 94249, 1090~GE Amsterdam, The Netherlands\\
$^4$ Department of Physics and Astronomy, Stony Brook University, Stony Brook, NY, 11794, USA}

\begin{document}

\date{Accepted 00 XXXX 00. Received 00 XXXX 00; in original form 00 XXXX 00}

\pagerange{\pageref{firstpage}--\pageref{lastpage}} \pubyear{}

\maketitle

\label{firstpage}

\begin{abstract}

Population synthesis studies constitute a powerful method to reconstruct the birth distribution of periods and magnetic fields of the pulsar population. When this method is applied to populations in different wavelengths, it can break the degeneracy in the inferred properties of initial distributions that arises from single-band studies. In this context, 
we extend previous works to include $X$-ray thermal emitting pulsars 
within the same evolutionary model as radio-pulsars. We find that the cumulative distribution of the number 
of X-ray pulsars can be well reproduced by several models that, simultaneously, reproduce the characteristics of the
radio-pulsar distribution. However, even considering the most favourable magneto-thermal evolution models with fast field decay, 
log-normal distributions of the initial magnetic field over-predict the number of visible sources with periods longer than $12$ s. 
We then show that the problem can be solved with different distributions of magnetic field, such as a truncated log-normal distribution, 
or a binormal distribution with two distinct populations. We use the observational lack of isolated NSs with spin periods
$P>12$ s to establish an upper limit to the fraction of magnetars born with $B > 10^{15}$ G (less than 1\%).
As future detections keep increasing the magnetar and high-B pulsar statistics, our approach can be used to establish a severe 
constraint on the maximum magnetic field at 
birth of NSs.

\end{abstract}


\section{Introduction}

The last couple of decades have seen tremendous advances in our
understanding of the properties and evolutionary paths of neutron
stars (NSs), largely driven by multi-wavelength detection.
Despite this progress, however, several puzzles still remain.  In
particular, the apparent diversity of the observational manifestations
of NSs, which has led over the years to several subclassifications
(i.e. rotation-powered pulsars, dim isolated NSs, Anomalous X-ray
Pulsars, Soft-Gamma Ray Repeaters, Central Compact Objects, among
others) has highlighted the need of understanding possible
evolutionary links among different classes, as well as the role played
by the birth properties of the NSs. 

Studying the properties of NSs as a whole is best done via Monte Carlo
simulations aimed at reproducing the galactic population.
Since the fraction of pulsars that can be detected close to their
birth constitutes a negligible fraction of the total sample, population synthesis studies
(PS in the following) generally use the present-day observed properties of pulsars,
together with some assumptions about their time evolution, to
reconstruct the birth distribution of periods and magnetic fields for
the pulsar population.

PS modeling of NSs has a long history, and especially so in the radio band,
where the largest sample of pulsars has been detected
 (e.g. \citealt{Gunn1970,Phinney1981,Lyne1985,Stollman1987,Emmering1989,Narayan1990,Lorimer1993,
Hartman1997,Cordes1998,ACC,Vranesevic2004,Faucher,Ferrario2006}) and in some cases in other
wavelengths \citep{Gon2004,Pierbattista}.
The efforts put over the years into this area of research stem from the
fact that these studies allow to constrain the birth properties of NSs,
which are in turn intimately related to the physical processes
occurring during the supernova (SN) explosion and in the proto-NS.  As
such, they bear crucial information on the physics of core-collapse
SNe, in which most NSs are believed to be formed.  

The energy reservoir powering Rotation Powered Pulsars (RPPs) is provided by the loss of rotational energy. 
RPPs are mostly seen in the radio band ($\sim 2200$ objects), and we can detect more than one hundred of them also in the
$X$-ray and/or $\gamma$-ray frequency range.
Part of the rotational energy loss is converted into non-thermal synchrotron and curvature radiation by acceleration of charged particles, which are either extracted from the surface or created by pair production. 
According to the classical scenario, such acceleration takes place in some regions, called gaps, where a cascade of pairs is supported by the same high-energy photons emitted by the particles, and is thought to produce the coherent radio emission we observe. 
While radio emission is likely produced above the polar cap \citep{sturrock71,ruderman75}, high-energy photons are thought to originate in the outer magnetosphere \citep{cheng86,romani96,muslimov03} or in the wind region \citep{petri12}. 
Of particular interest is the recent FIDO model \citep{Kalapo2014}, according to which the gamma-ray emission is produced in regions near the equatorial current sheet. These models can fit the observed values of the radio-lags and reproduce the observed light curve phenomenology and the GeV photon cut-off energies. 

The majority of PS studies in the literature have assumed that
the magnetic field is fixed at its value at birth, or have adopted
simple, parametrized expressions for its decay.  
Recently, a large theoretical effort has been devoted to model the magnetic field evolution with 
magnetothermal simulations which follow the coupled evolution of the temperature and the magnetic field in the
NS interior \citep{Vigano2013}. We will use results from these simulations to model
the magnetic field evolution.

\cite{Gullon2014} (paper I, hereafter) have revisited PS modeling of isolated
radio-pulsars incorporating the $B$-field evolution from the
state-of-the-art magneto-thermal evolution models, 
together with recent results on the evolution of the angle between the magnetic and
rotational axes from magnetospheric simulations.  
In an era where the availability of high energy satellites has enlarged considerably our understanding of the NS population, 
unvealing a variety of observational manifestations of NSs, it is then timely to include these further constraints
 in the global PS study of the NS population. 
A first attempt in this direction, including part of the X-ray emitting NS population and magnetic field decay models, was
performed by \cite{PSB_Popov}, with the specific goal of uncovering
links among the apparently different NS manifestations. 

In line with recent theoretical efforts aimed at understanding the various observational manifestations of NSs, with a unifying
evolution model \citep{PernaPons, PonsPerna,Vigano2013}, 
our goal in this work is to combine previous studies of the radio-pulsar population (paper I) with the study of the population of
$X$-ray pulsars. 
For young and middle-age NSs ($\lesssim 1$ Myr), the thermal emission comes from the internal heat which is gradually released 
or, in the case of NSs with large magnetic fields ($B\gtrsim 10^{14}$ G),  by Joule dissipation of the electrical currents circulating 
in the NS  crust which enhances the thermal luminosity.
We do not include in this study non-thermal emission 
powered by losses of rotational energy, mainly because of the lack of detailed models for the physical mechanisms.
Predictions for the non-thermal luminosities are based on phenomenological correlations 
with the rotational energy losses, which include more free parameters that are not linked to physical properties. 
On the contrary, independent, physically motivated cooling models provide the information linking the model parameters with
the thermal emission of NSs. We hence focus on this population of NSs and
leave the study of rotation powered, non-thermal emission ($X$-ray and $\gamma$-ray pulsars) for future works.

The article is organized as follows: 
In \S~\ref{sec:mod} we summarise the main results of paper I about the radio-pulsar population and we describe the assumptions made to model the X-ray thermal radiation. In  \S~\ref{sec:results} we discuss the results of our analysis and finally, 
in \S~\ref{sec:conclusions} we present our conclusions and final remarks.




\begin{table*}
\begin{center}
\begin{tabular}{|l|cccccccccccr}
\hline
\hline
Model	&Decay&	Envelope     & $\mu_{B_0}$ 	& $\sigma_{B_0}$ 	& $\mu_{P_0}$  &  $\sigma_{P_0}$ & $\alpha$ & $n_{\rm br}$ & ${\cal{D}}$\\
      &	&	 & $\log{B}$ [G] 	& $\log{B}$ [G] & [s]  	&  [s] & &  [century$^{-1}$] & & & \\
\hline
A	& No decay	&	Heavy elements 		&	$12.65$ 	& $0.50$ 		& $0.38$ 		& $0.35$ 		& $0.50$ & $5.70 \pm 0.16$ & $0.087 \pm 0.017$  &  \\
B	& Slow 	&	Heavy elements   		&	$13.04$ 	& $0.55$ 		& $0.23$ 		& $0.32$ 		& $0.44$ & $2.63 \pm 0.05$ & $0.082 \pm 0.010$  &  \\
C	& Fast      &	Heavy elements   		&	$13.20$ 	& $0.72$ 		& $0.37$ 		& $0.33$ 		& $0.41$ & $3.82 \pm 0.14$ & $0.106 \pm 0.011$  &  \\
D	& Slow      &	Light elements      	&	$12.99$ 	& $0.56$ 		& $0.16$ 		& $0.31$ 		& $0.43$ & $2.63 \pm 0.06$ & $0.078 \pm 0.009$  &  \\
E	& Medium    &	Light elements     	&	$13.13$ 	& $0.68$ 		& $0.32$ 		& $0.19$ 		& $0.44$ & $3.16 \pm 0.11$ & $0.083 \pm 0.007$ &  \\
F	& Slow (with toroidal field) &	Light elements    & $12.99$	& $0.59$ & $0.21$	 	& $0.32$		& $0.44$ & $2.95 \pm 0.09$ & $0.083 \pm 0.008$ & \\

\hline
\hline
\end{tabular}
\end{center}
\caption{Optimal parameters for the radio-pulsar population for each magnetothermal evolution model considered in this work.
Columns $4$ to $8$ indicate the values of the free parameters that minimize the ${\cal{D}}$-value 
computed through a 2D KS test.
Column $9$ shows the corresponding birthrate, and column 10 shows the average ${\cal{D}}$-value of 10 realizations and
its standard deviation $\sigma$.}

\label{tab:par_rad}
\end{table*}


\begin{figure}
\begin{center}
\includegraphics[width=7.2cm]{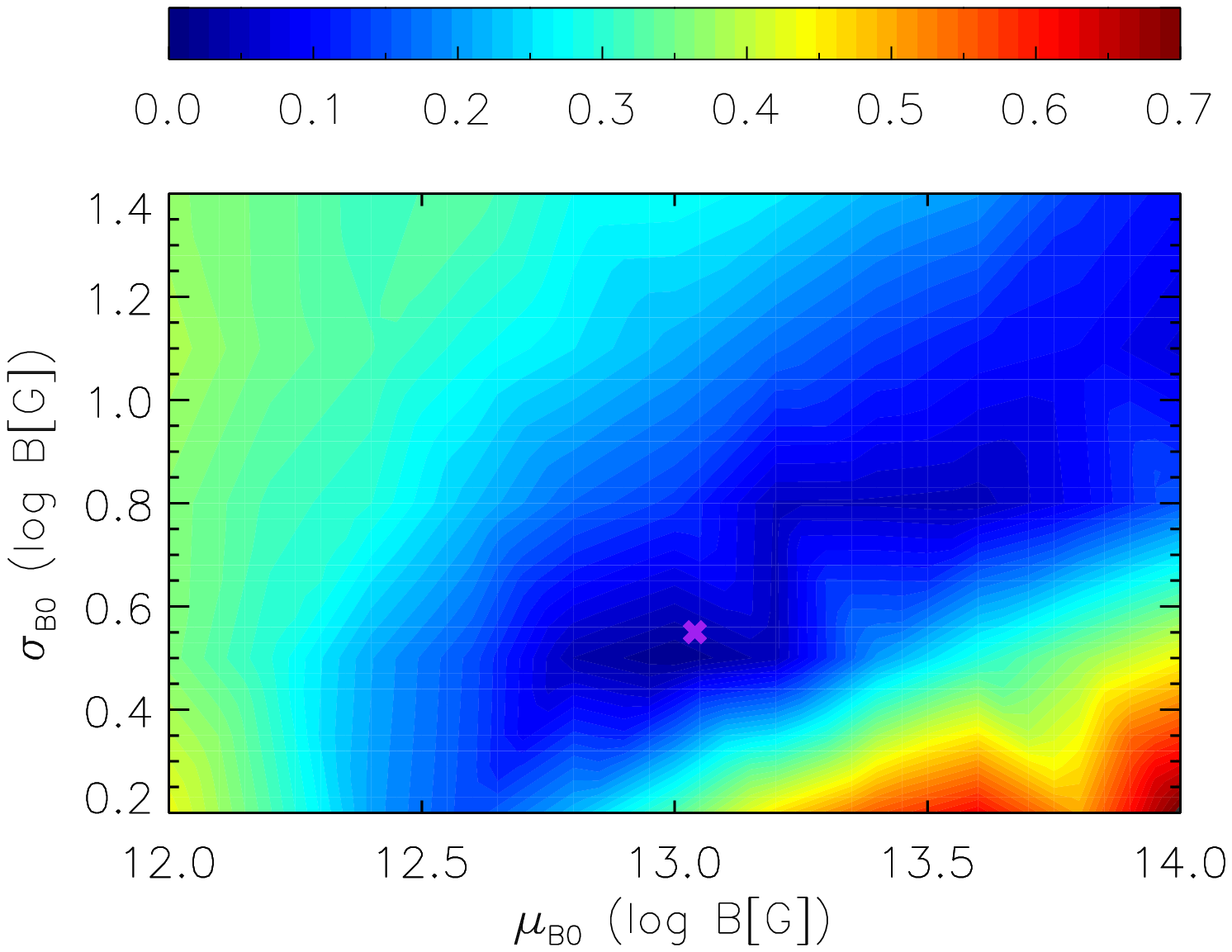}
\includegraphics[width=7.2cm]{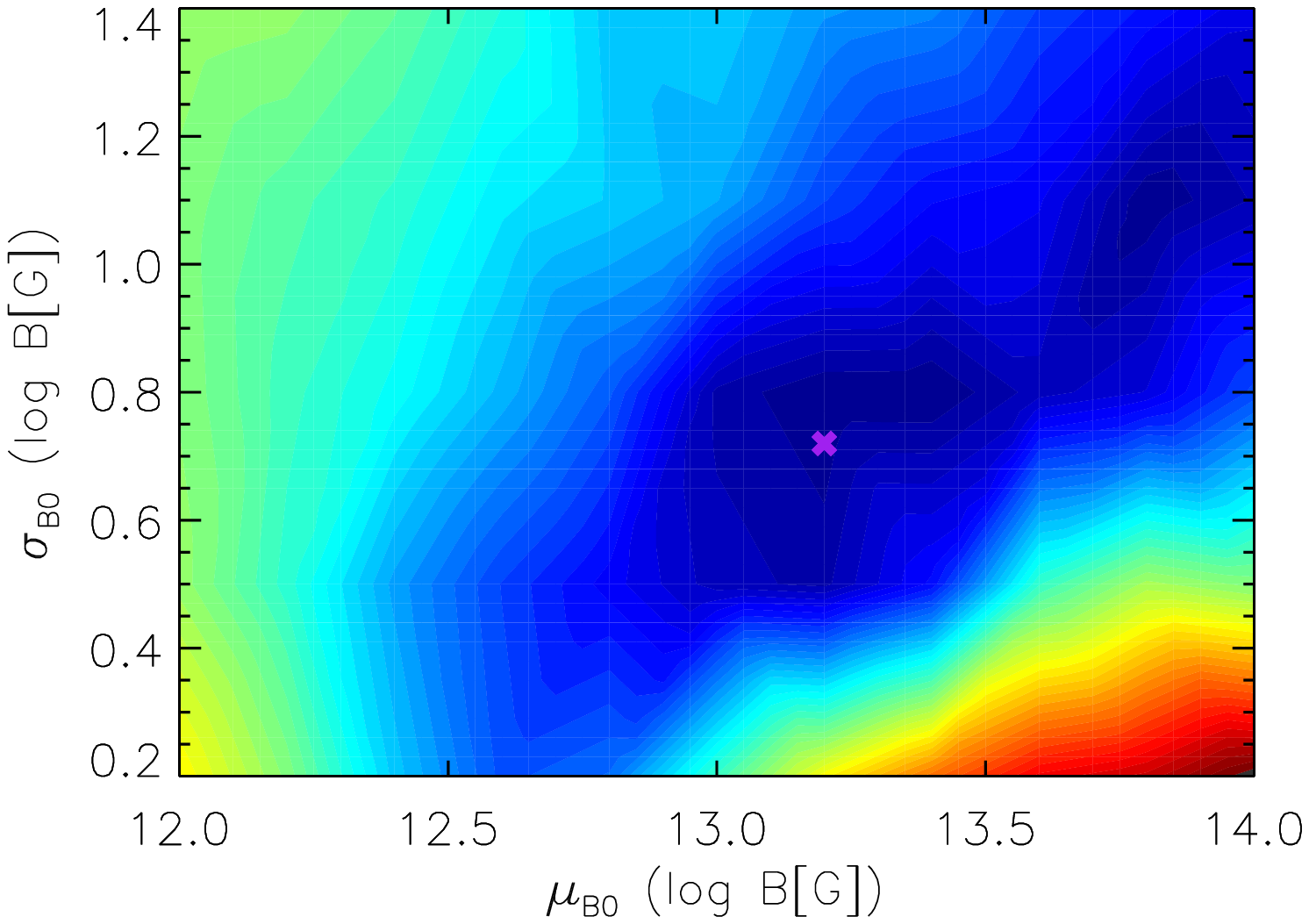}
\includegraphics[width=7.2cm]{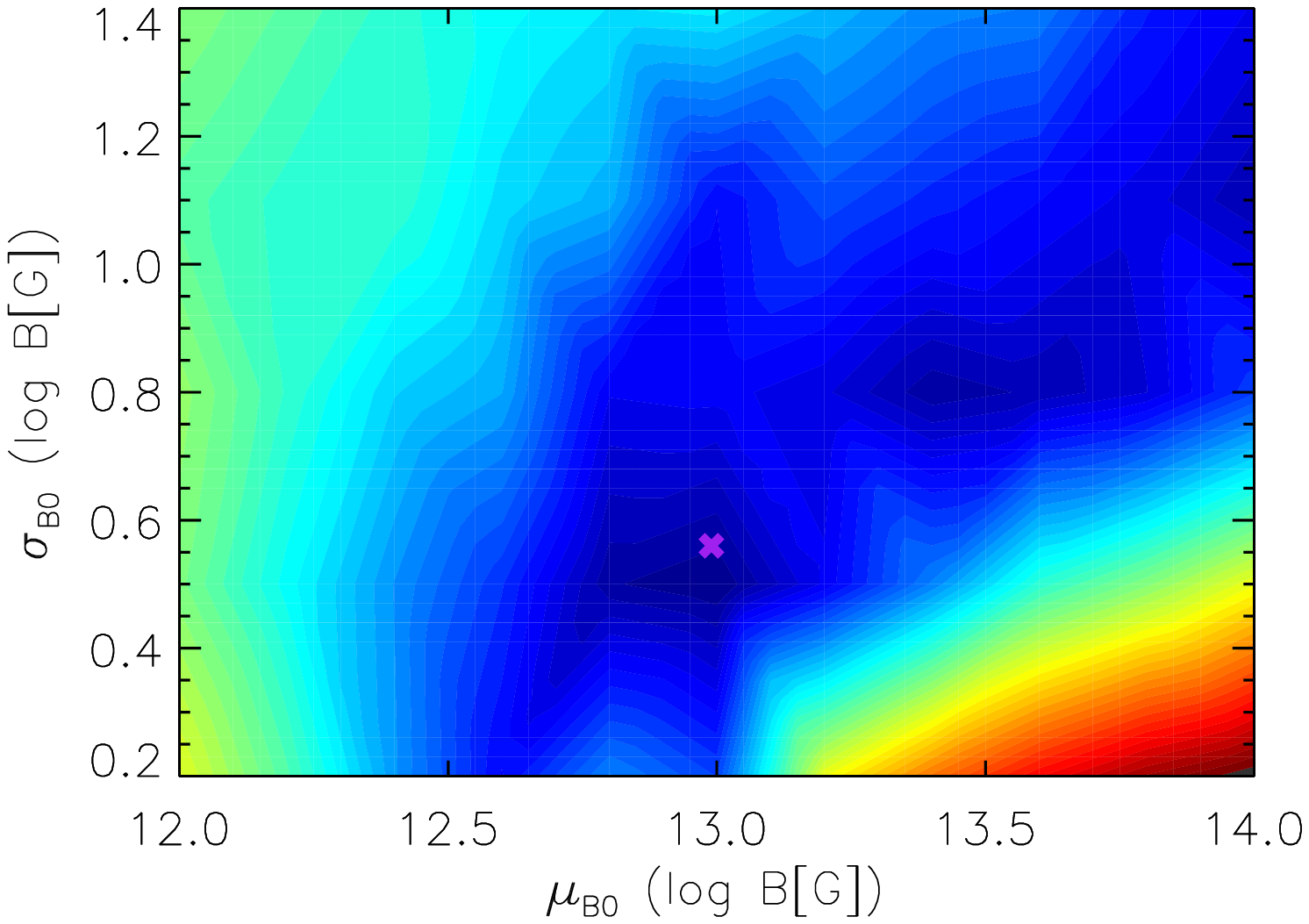}
\end{center}
\caption{Contour plot of the ${\cal{D}}$-value computed through a 2D KS-test in the $\mu_{B_0}$ - $\sigma_{B_0}$ plane 
(a 2D cut of the general 5D parameter space), for models B (top), C (middle) and D (bottom).
The purple crosses indicate the minima found by the simulated annealing procedure in each case, and listed in Table~\ref{tab:par_rad}.}
\label{fig:b0}
\end{figure}


\section{Multi-wavelength population synthesis}\label{sec:mod}

From an observational point of view, there is an important qualitative difference, regarding detectability biases, 
between the radio band and the high-energy bands. In the former, relatively complete surveys allow to quantitatively implement the instrumental detectability in the PS study. In $X$-rays, instead, the sample is not complete, and the sky coverage is inhomogeneous. Moreover, the detectability depends on non-trivial issues, such as the presence of a counterpart with known ephemerides in other wavelengths, or whether or not it has been associated to a transient $X$-ray outburst. 
This implies that the potential number of visible NSs  above a certain 
threshold flux should be larger than the actual number of detected sources.

In $X$-rays we have to account for the effect of absorption by the interstellar medium.
The hydrogen column density, $N_{H}$, for each synthetic pulsar is:  
\begin{equation}
N_{H} = \int n_{H}\,dl
\end{equation}
where $n_{H}$ is the local hydrogen number density, whose spatial distribution is estimated following \cite{nh}, and the integral is performed along the line of sight $l$.
The energy-dependent absorption is calculated following the models of \cite{phabs},
which include photoelectric cross sections of $17$ elements in the $[0.1-10]$ keV range. This allows us to calculate the detected (absorbed) flux on Earth $S_{X}^{\rm abs}$, assuming that radiation is isotropically distributed.

\subsection{The radio pulsar population}

For completeness, we begin our discussion by reviewing the main results of paper I, to which we refer for technical details.
An outline of the procedure to generate a NS population is sketched in the following way:

\begin{itemize}

\item A large number of NSs with uniformly distributed random ages are generated in the framework of a given magneto-thermal 
evolution model (we explore six different models), which determines the evolution of the magnetic field.
The birth location distribution, kick velocity, and evolution in the galactic gravitational potential are described in paper I.

\item
Gaussian distributions are assumed for both the initial spin period $P_0$ and the logarithm of the initial magnetic field, $B_0$.
The central values ($\mu_{B_0}$, $\mu_{P_0}$) and widths ($\sigma_{B_0}$, $\sigma_{P_0}$) are left as free parameters.
The radio-luminosity (at 1400 MHz) obeys the following equation:
\begin{equation}
L_{1400} = L_{0} \, 10^{L_{\rm corr}} \, (P^{-3} \dot{P})^{\alpha} ~,
\label{radiolum}
\end{equation}
where $P$ is the spin period, $\dot{P}$ is the period time derivative, $L_{0} = 5.69 \times 10^6$ mJy kpc$^2$ and $L_{\rm corr}$ is a random correction chosen from a zero-centered gaussian of 
$\sigma = 0.8$ (as in \citealt{Faucher}), to take into account the large observed dispersion. 
The index $\alpha$ is a free parameter.

\item The NSs are classified as {\it radio-visible} according to the same flux and beam visibility criteria of the radio survey used in paper I. The generation of pulsars stops when the number of radio-visible pulsars is equal to the number of radio pulsars in the considered 
observational sample. This fixes our normalisation for the total number of sources or, equivalently, the birth rate $n_{\rm br}$ for each synthetic population.

\item The set of free parameters ($\mu_{B_0}$, $\sigma_{B_0}$, $\mu_{P_0}$, $\sigma_{P_0}$ and $\alpha$)
is determined for each model by minimizing the associated ${\cal{D}}$-value of a generalized two-dimensional Kolmogorov-Smirnov test.
The ${\cal{D}}$ statistic is calculated from the four natural quadrants that define each point in the $P-\dot{P}$ plane of the observed sample.
The fraction of data from both samples (observed and simulated) in the radio band, is calculated on each quadrant, and
the ${\cal{D}}$-value is defined as the maximum difference between observed and simulated fractions over all points and quadrants
(for a more technical description we refer the reader to \citealt{KS2D,numericalrecipes}).
 
\end{itemize}

The results for the 6 evolutionary models studied are summarized in Table~\ref{tab:par_rad}.
Here we considered three models with iron envelopes (A, B and C, already presented in paper I), differing in 
the timescale of the magnetic field decay. We have also used three new models (D, E and F) with light element envelopes.
The last one also has an additional strong toroidal magnetic field.
Although the envelope composition barely affects the evolution of rotational properties ($P$ and $\dot{P}$), it has a relevant
impact on the thermal luminosity (see next section).

In general, we find that the differences in the initial period distribution between different models are not significant.
Indeed, any broad distribution of initial periods in 
the range $0 < P_0 < 0.5$ s can be reconciled with the data, but narrow distributions peaked at short spin periods 
($P_0 < 100$ ms) are ruled out. 
The ${\cal{D}}-$value is more dependent on the initial magnetic field distribution, but this is not strongly constrained
because its central value and width are correlated showing an interval of possible solutions: the larger the central value, 
the wider the distribution.
This holds true for models with and without field decay, and with different assumptions for the magnetospheric torque.
This correlation is visible in Fig.~\ref{fig:b0}, where we show contour plots of the ${\cal{D}}$-value in the 
$\mu_{B_0}-\sigma_{B_0}$ plane. We note this is a 2D plot of a section of a 5D parameter space. 
For conciseness, we only show plots for models B,C, and D (other models look very similar).

The dark blue regions, where the ${\cal{D}}$-value is low, 
define the range of allowed parameters. The purple crosses mark the different solutions for which explicit values for the parameters
have been given in Table~\ref{tab:par_rad}.
The main conclusion from the analysis of only the radio pulsar population is that, independently of the underlying physical model,  
one can find acceptable parameter sets to fit the radio population reasonably well. Thus the radio-pulsar population alone
cannot help much to discriminate among different evolutionary models. The next question is
whether or not the analysis of pulsar distributions in other bands can break the degeneracy.

\begin{figure*}
\begin{center}
\includegraphics[width=5.6cm]{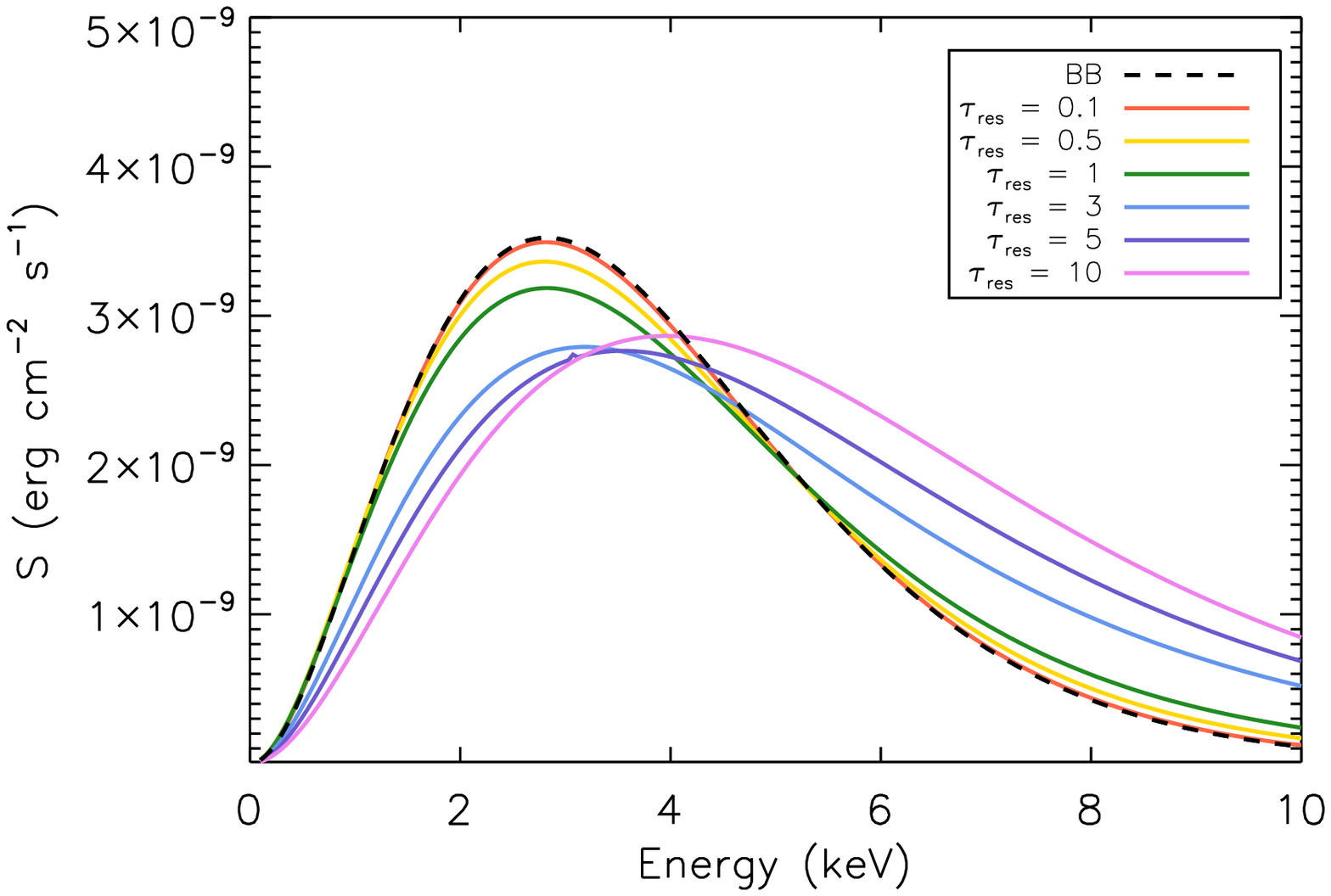}
\includegraphics[width=5.6cm]{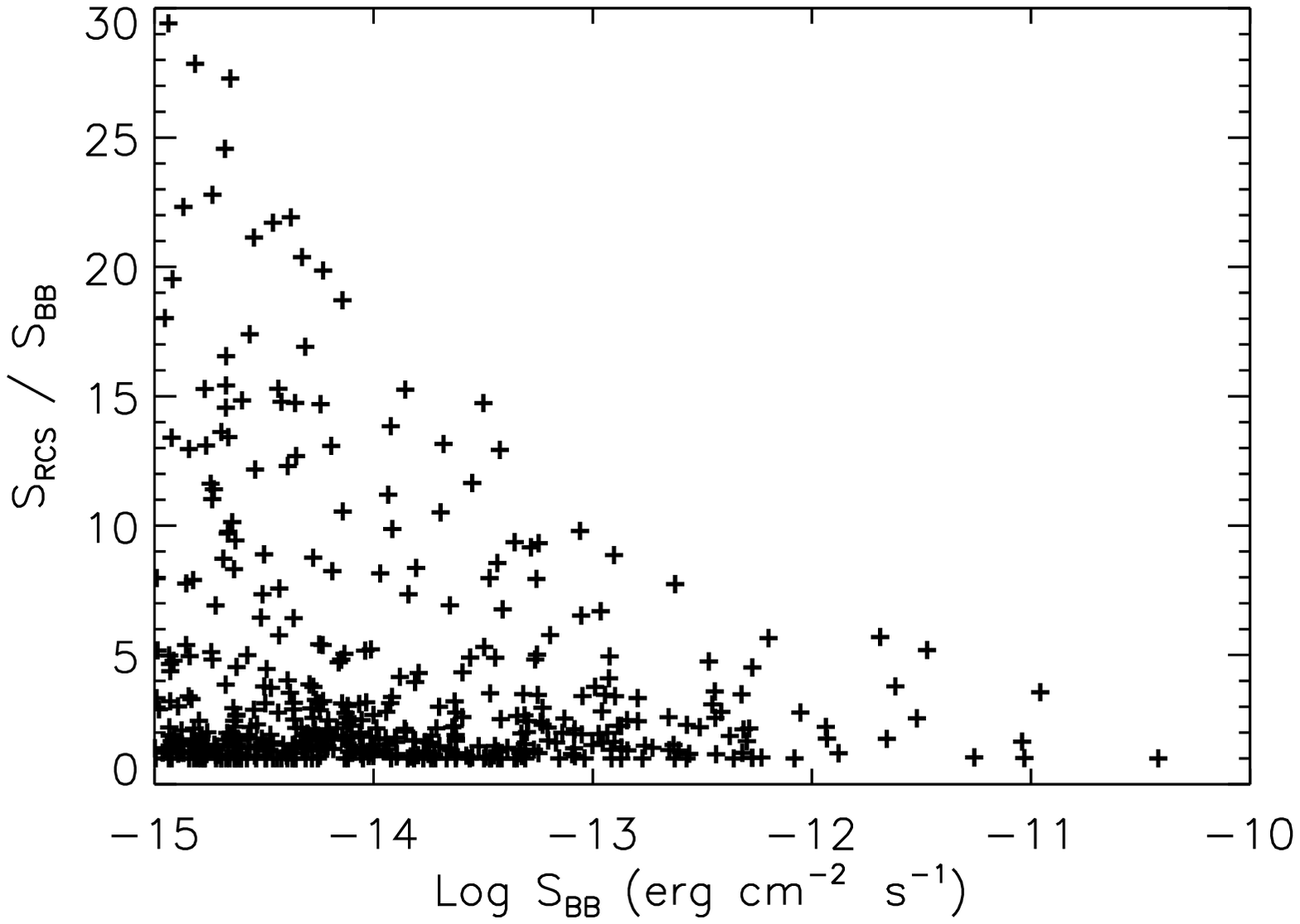}
\includegraphics[width=5.6cm]{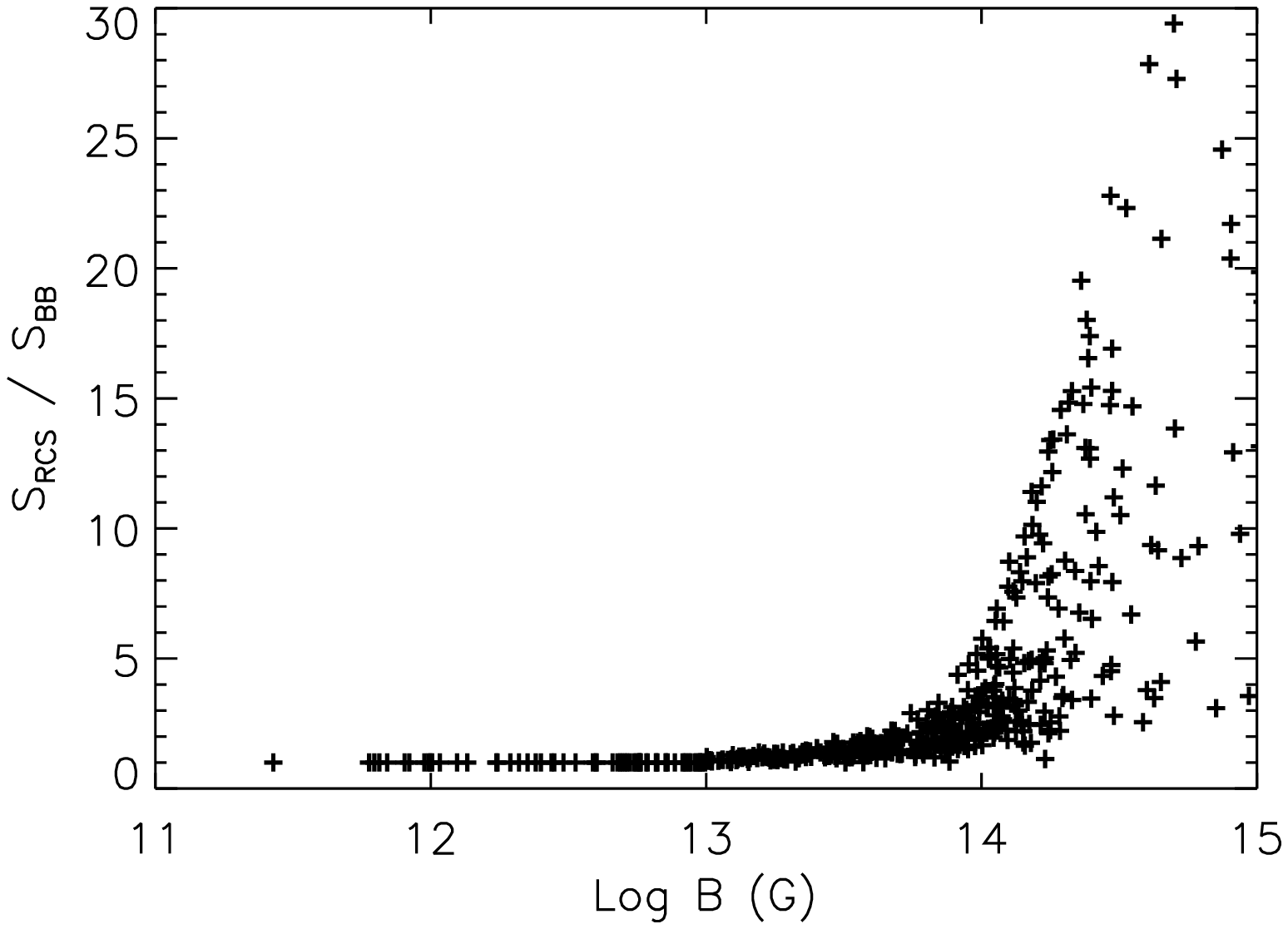}
\end{center}
\caption{{\em Left}: Blackbody spectrum with $kT=$ 1 keV (dashed line), and reprocessed spectra using the RCS model with $\beta_T = 0.3$ and $\tau_{0}=0.1,0.5,1,3,5,10$ (solid color lines). Unabsorbed fluxes with a typical galactic value of $N_{H} \sim 3 \times 10^{22}$ cm$^{-2}$ are plotted. {\em Middle}: Ratio between the absorbed flux of pure blackbody model to that of the RCS model as a function of the blackbody flux $S_{BB}$. {\em Right}: same as the middle panel, as a function of the
logarithm of the magnetic field.
The results show only the NSs with detectable thermal emission in the 0.1-10 keV band for model E,  with fluxes  above $S_{Xmin} = 10^{-14}$ erg s$^{-1}$ cm$^{-2}$.}
\label{fig:rcs}
\end{figure*}

\subsection{Thermally emitting NSs}

\subsubsection{The observational sample}

We consider all NSs for which the thermal
emission is dominant, or clearly contributing to the total power in the soft X-ray band, and whose origin is attributed
to residual cooling from birth.
The selected sample we use has been described in \cite{Vigano2013} and it is periodically updated in our website\footnote{\url{www.neutronstarcooling.info}}, includes thirteen closeby RPPs (only those with a clear thermal component), seven X-ray dim Isolated NSs (XINSs), and seventeen magnetars.

The XINSs are a small group of nearby (within about 200-300 parsecs), thermally emitting, isolated NSs.
They are radio quiet but relatively bright in the X-ray band, and
five of them have periods similar to magnetars. Moreover,
their inferred magnetic fields are between the upper end of the radio
pulsar population and the magnetar candidates. At variance with the
majority of X-ray emitting pulsars, XINSs spectra are rather
perfect blackbodies (kT$\sim$0.1 keV, occasionally with some broad
lines), which make them the key objects for research
on the NS equation of state, and the cooling properties of
the NS crust and core.

The ``magnetars" \citep{olausen14} are a small group of X-ray pulsars with spin periods
between $0.3-12$s, whose properties cannot be explained within
the common scenario for pulsars. 
Their inferred magnetic fields are $B \sim10^{14-15}$\,G. The decay and instabilities of such large magnetic fields are thought to power their emission \citep{duncan92,thompson93}. Their powerful persistent X-ray emission (much larger than the rotational energy loss)
is well modelled by a blackbody with $kT \sim 0.2-0.8$\,keV with an additional power-law 
component ($\Gamma\sim2-4$), usually interpreted as the product of Resonant Compton Scattering (RCS) of the seed thermal 
photons in the dense magnetosphere \citep{thompson02}. Hence even this apparent non-thermal component has a thermal origin. 
Different physical models have succeeded to reproduce the effect of the RCS to fit the magnetars X-ray spectra \citep{rea08,zane09,beloborodov13}.
Given the transient nature of these objects (they show flares and sometimes long radiative X-ray outbursts), we have 
used in this work only X-ray luminosities derived during their
quiescent state (available for the 17 selected objects).

\subsubsection{Magneto-thermal evolution and emission model}

In NSs endowed with strong magnetic fields, the evolution of the internal and surface temperatures and the magnetic field are 
closely linked. The thermal luminosity is strongly dependent on the time-dependent magnetic field strength and topology, while
the evolution of the magnetic field depends on the local values of the magnetic diffusivity and the electron relaxation time, 
which strongly depend on the temperature. \cite{Vigano2013} performed 2D magneto-thermal simulations of NSs, and studied 
the evolution of magnetic field strength and thermal luminosity as a function of time. 
Using this code, we produce a number of tables with the effective temperature and value of the dipolar magnetic 
field as a function of the age of the NS.
Among the many model parameters (initial magnetic field topology, mass, radius, envelope composition, etc), we will 
focus on the parameters that more strongly influence the observable quantities (luminosity, $P$, $\dot{P}$) at a given age. 
These are the envelope composition (light vs. heavy elements), the initial magnetic field strength at the pole, $B_0$, 
and the impurity parameter in the inner-crust $Q_{imp}^{ic}$, a measure of the average charge distribution of the nuclei present
in the crust. We refer to \cite{Vigano2013} for more details.

The magnetic field strength determines the rotational evolution of the NS. In addition, in highly magnetised NSs, the additional
energy reservoir provided by the  magnetic field dissipation may further increase the surface temperature, and extend the
duration of the stage in which neutron stars are bright enough to be seen as X-ray pulsars.
$Q_{imp}^{ic}$ was found to be a crucial parameter to understand the clustering of spin periods in isolated X-ray pulsars 
and why there is an upper limit of isolated X-ray pulsars at about 12 s (\citealt{Pons2013}).
The composition of the envelope is also of particular relevance.
Generally speaking, low-B NSs with light element envelopes have significantly larger temperatures
and luminosities during the first $\approx 10^5$ yr of their life, and lower temperature/luminosity afterwards, once they enter the 
photon cooling era \citep{Page2004,Yakovlev2004,Potekhin2007,Page2011}. 

These three parameters characterize the evolution of the magnetic field strength and temperature as the star ages, 
and therefore the trajectories that NSs follow in the $P$-$\dot{P}$ diagram. Having all these tabulated quantities computed via
simulations, for each star generated in the synthetic sample we calculate the present value of magnetic field, $P$, $\dot{P}$, temperature and luminosity. We assume blackbody emission, and integrate over the whole star surface to
calculate the luminosity. For NSs with high magnetic fields,
we also add the effect of resonant Compton scattering (see next subsection) in the spectrum. 
Among {\it all NSs generated} (not only those
seen as radio-pulsars), we select those for which the absorbed X-ray flux at the Earth location is above 
$10^{-15}$ erg s$^{-1}$ cm$^{-2}$, as potentially visible X-ray pulsars (for present or future instruments).

\subsubsection{Resonant Compton Scattering}

The thermal emission is usually assumed to be blackbody-like. However, it is well known that magnetar spectra show a non-thermal tail, whose origin cannot be attributed to the rotational energy of the star. The most popular interpretation is that magnetospheric 
plasma boosts up thermal photons coming from the surface.
For these reasons, it is important to account for the possible effects that spectral distortions in the form of a hard tail would have on the detectability of NSs in X-rays. 

We adopt a simplified model for the resonant Compton scattering (RCS, \citealt{lyutikov06}) of seed photons (assumed to have
a pure blackbody spectral energy distribution) in NS magnetospheres, and successfully applied to model spectra of several magnetars \citep{rea08}. Although it is a 1D, plane-parallel model, far from describing the non-trivial interaction between plasma and radiation in a realistic 3D magnetosphere, it 
effectively fits X-ray data. RCS models are parametrized by two parameters: the  resonant optical depth along a ray
$\tau_{res}$, which is related to the plasma density, and the thermal velocity of electrons in units of the speed of light, $\beta_{T}$
(see \citealt{lyutikov06} for more details).

In Fig.~\ref{fig:rcs} (left) we show the comparison between a blackbody of $kT= 1$ keV, and the spectra predicted by the RCS model for $\beta_T=0.3$ and different values of $\tau_{res}$. 
We show the spectra as they would be seen when absorbed by the  interstellar medium with $N_H=3\times 10^{22}$ cm$^{-2}$, a typical
value for magnetars in the galaxy. Since the interstellar absorption is important below $\sim 1$ keV, an efficient RCS could enhance the NSs detectability, because it promotes 
low energy photons to higher energies.

\cite{rea08} fit the X-ray spectra in quiescence of several magnetars. We found a correlation between 
$B$ and the RCS parameters $\beta_T$ and $\tau_{res}$, which can be reproduced by the following analytical fits:
   \begin{equation}
     \beta_T = \left\{
	       \begin{array}{ll}
		 0.001      & \mathrm{if\ } \log{B} \le 13 \\
		 0.3     & \mathrm{if\ } \log{B} > 13
	       \end{array}
	     \right.
   \end{equation}

   \begin{equation}
     \tau_{res} = \left\{
	       \begin{array}{ll}
		 0.001      & \mathrm{if\ } \log{B} \le 13 \\
		 \frac{B}{10^{14} {\mathrm G}}     & \mathrm{if\ } \log{B} > 13~.
	       \end{array}
	     \right.
   \end{equation}

We implement this analytical fit to modify the spectra of the randomly generated NSs in the population synthesis code. RCS does 
not change the original blackbody spectrum in NSs with low magnetic field, but it generates a hard tail in magnetars. 
In order to show the effect of the RCS in a typical PS realisation, in the middle panel of Fig.~\ref{fig:rcs2}
we show the ratio of the RCS flux to the pure blackbody flux in the $[0.1,10]$ keV band, as a function of the blackbody flux.
This figure clearly shows the main effect of the RCS correction: a flux enhancement by a factor of a few
(up to a factor of 20) is applied to a significant number of sources, which has a visible imprint in the statistics of the number of detectable NSs. This effect is more evident for the dimmest sources: since RCS affects the high-energy tail,  we can only obtain a very large enhancement for sources with a large $N_H$, 
and these are always dim. The right panel of Fig.~\ref{fig:rcs} shows the same ratio as a function of the initial field strength. It shows that only for the highest fields, in the magnetar range, a significant flux enhancement in the [0.1-10] keV band is produced. 

\begin{figure}
\begin{center}
\includegraphics[width=8.5cm]{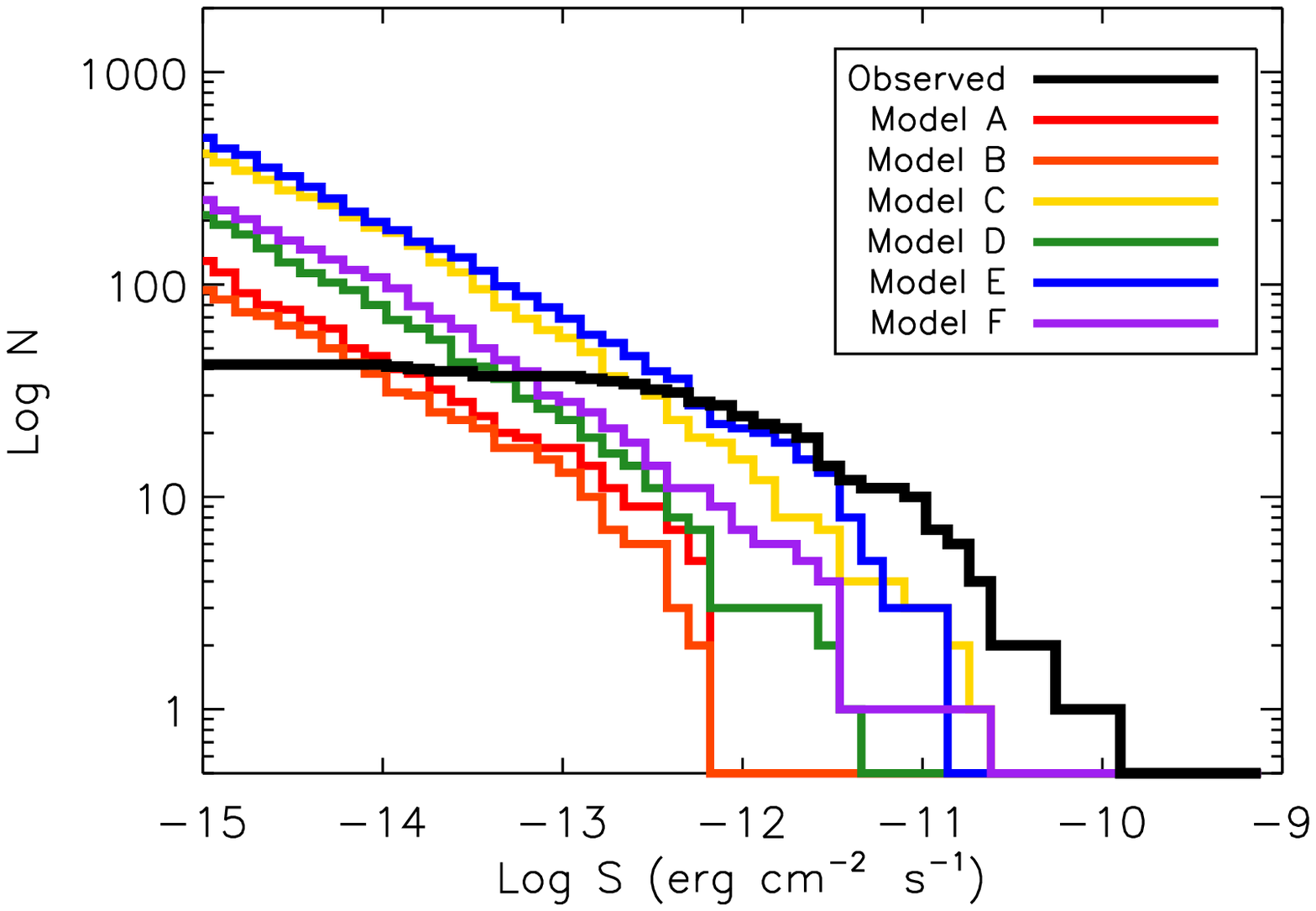}
\includegraphics[width=8.5cm]{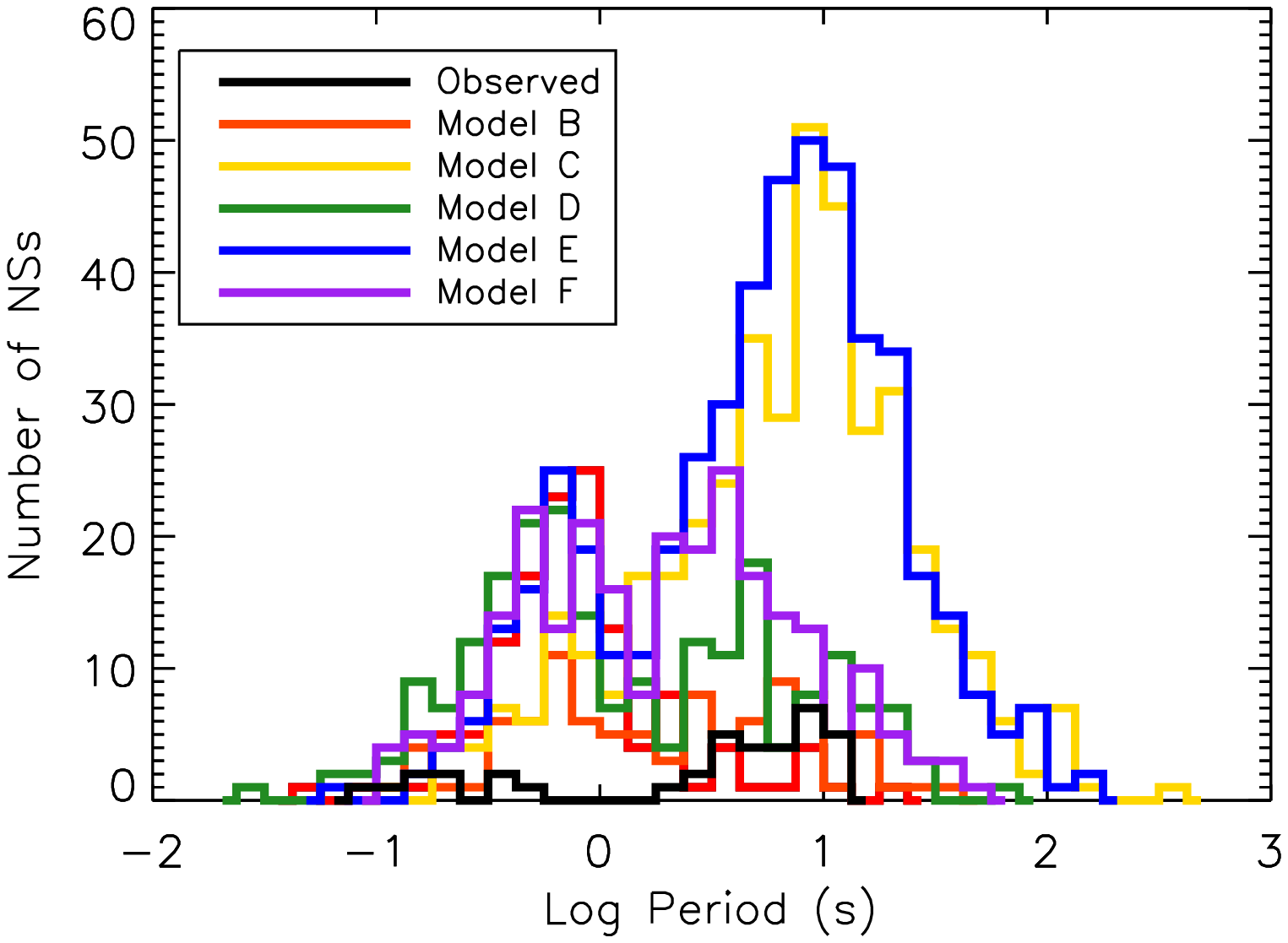}
\end{center}
\caption{Log N - Log $S$ diagrams ({\it top}) and period distributions ({\it bottom}) for the thermally emitting X-ray pulsars for models 
A to F and the parameters listed in Table \ref{tab:par_rad}.
In the bottom panel we show the histogram of sources with $S_{Xmin} > 10^{-14}$ erg s$^{-1}$ cm$^{-2}$.
In colour lines we show the population synthesis results while the black line shows the observed sources.}
\label{fig:lognlogs_rad}
\end{figure}


\section{Results}\label{sec:results}

We begin by analyzing the solutions that best fit the radio-pulsar distribution (Table \ref{tab:par_rad}).
In Fig.~\ref{fig:lognlogs_rad} we compare the accumulated number of detected X-ray sources $N(S)$ with flux 
$S>S^{abs}_{X}$  (hereafter $\log{N}$-$\log{S}$ diagram) for each magnetothermal evolution model.
The sample of sources is not complete, since a full all-sky survey is 
not available for the present instruments, and 
many sources, in particular magnetars, have been discovered while they are in outburst. We can assume that we are actually
observing all of the few brightest sources, and we gradually miss more and more sources as the threshold flux decreases.
From a quick inspection of the $\log{N}$-$\log{S}$ diagrams, we can immediately conclude that none of the models of
Table 1 predicts a number of thermally emitting X-ray sources compatible with the observations. 
However, as shown in Fig.~\ref{fig:b0}, there is a quite large region in the 
parameter space with similar statistical significance for the ${\cal{D}}$-value. We can then explore whether other parameter 
combinations (always within the region allowed by the radio-pulsar population analysis) can improve the $\log{N}$-$\log{S}$
results. Since we obtain a systematic lack of bright sources for all models, we have to modify the initial distribution to include more
NSs born with high magnetic fields, which can be done if we move up-right in the dark blue region of Fig. \ref{fig:b0}.

\begin{figure}
\begin{center}
\includegraphics[width=8.5cm]{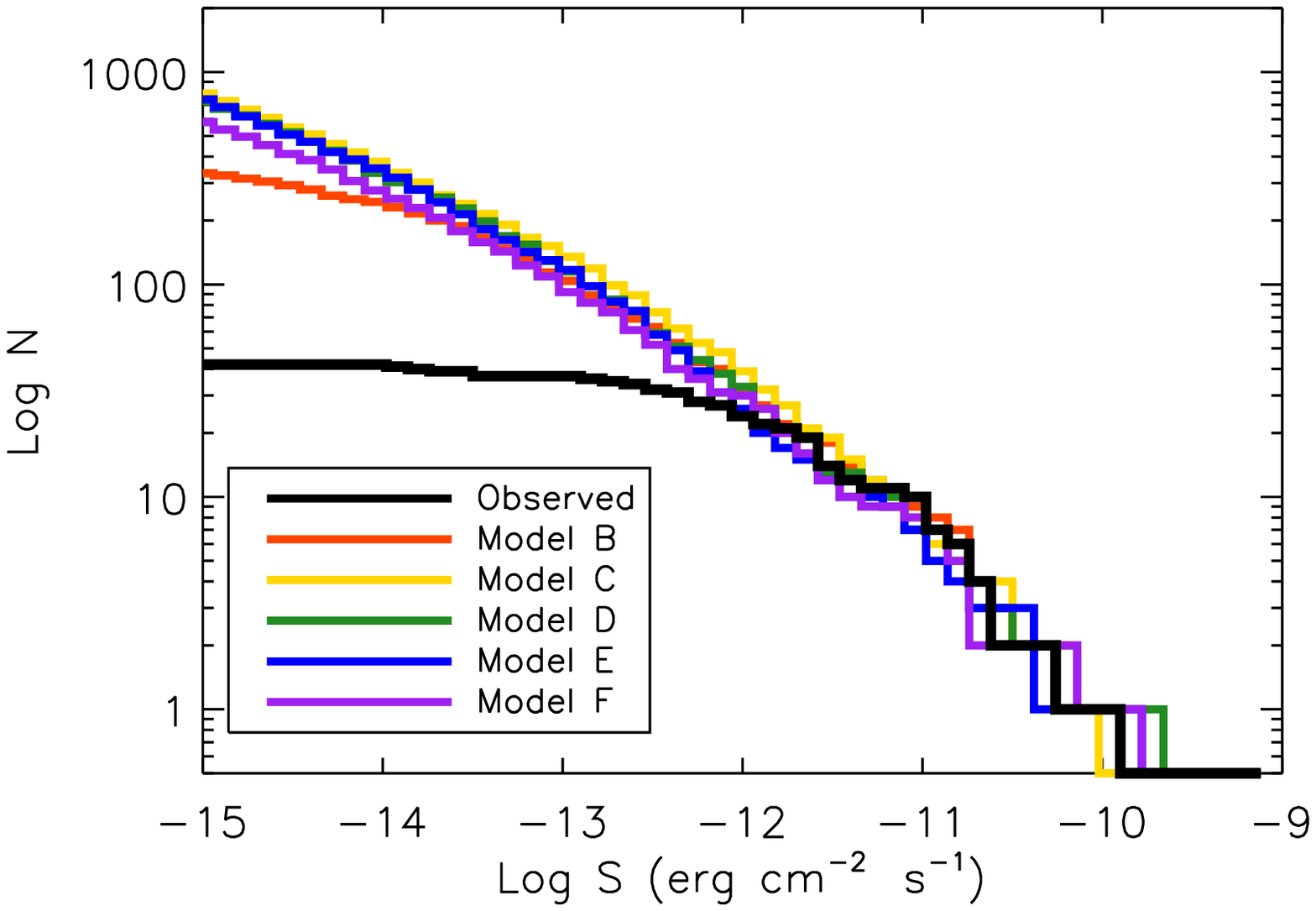}
\includegraphics[width=8.5cm]{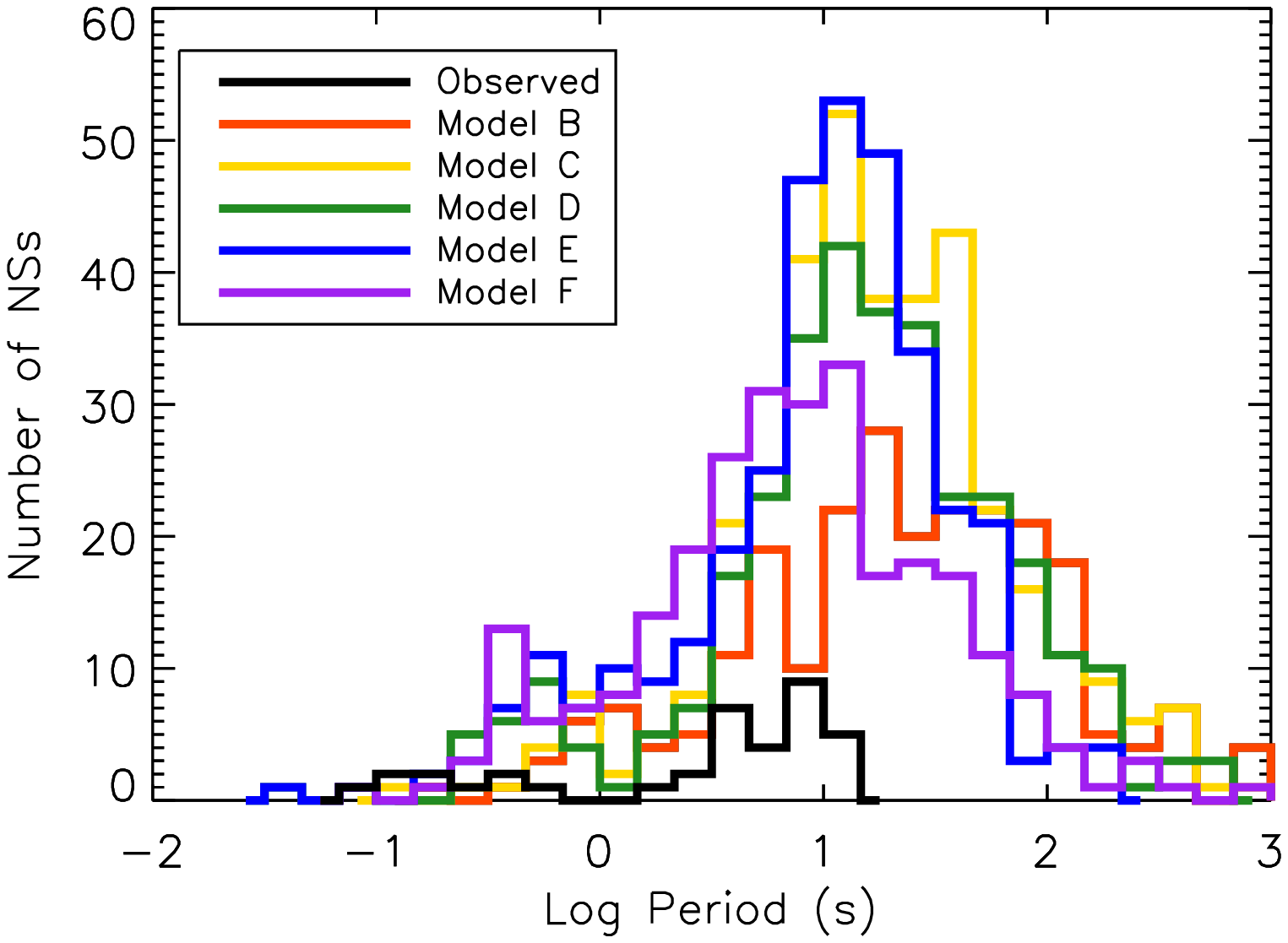}
\end{center}
\caption{Same as Fig.~\ref{fig:lognlogs_rad} for the parameters listed in Table \ref{tab:par_X}.}
\label{fig:X}
\end{figure}

We have proceeded as follows: first, we search for a pair of values of $\mu_{B_0}$, $\sigma_{B_0}$ in the dark blue region of Fig. \ref{fig:b0} than can reproduce the observed distribution of X-ray pulsars for fluxes above $\approx 10^{-12}$ erg s$^{-1}$ cm$^{-2}$ 
(the other 3 parameters do not affect the X-ray population). Once this is done, we fix the values of these two parameters and repeat the minimization procedure to find new values of $\mu_{P_0}$, $\sigma_{P_0}$, and $\alpha$ that best fit the radio-pulsar distribution.
In Table~\ref{tab:par_X} and Fig.~\ref{fig:X} (top) we show the results for the parameters that best reproduce the 
$\log{N}$-$\log{S}$ diagram for all models except model A, for which field decay was not included, and therefore 
the luminosity is not affected by the magnetic field. In other words, the standard cooling of a neutron star without including 
any additional heating by magnetic field decay cannot be reconciled with the X-ray data.
For the other models, we find acceptable solutions with similar statistical significance for the radio-pulsar fits. The $\cal{D}$-value
obtained with the 2D KS test shown in the last column is an average of 10 realizations, and we can see that they are compatible
with the results of Table 1 within 1- or 2-$\sigma$.

\begin{table*}
\begin{center}
\begin{tabular}{|l|ccccccccr}
\hline
\hline
Model & $\mu_{B_0}$ 	& $\sigma_{B_0}$ & $\mu_{P_0}$  &  $\sigma_{P_0}$ & $\alpha$ & $n_{\rm br}$     &	$\eta$	&	${\cal{D}}$	\\
      & $\log{B}$ [G] 	& $\log{B}$ [G]  & [s]  	  &  [s]            &          & [century$^{-1}$] &		      &			\\
\hline
B	& $13.34$ 		& $0.76$ & $0.30$ & $0.29$ & $0.44$ & $4.98 \pm 0.11$ & $3.1$ & $0.089 \pm 0.009$\\
C	& $13.36$ 		& $0.80$ & $0.29$ & $0.37$ & $0.44$ & $4.37 \pm 0.12$ & $4.0$ & $0.093 \pm 0.011$\\
D	& $13.25$ 		& $0.77$ & $0.28$ & $0.20$ & $0.43$ & $3.07 \pm 0.11$ & $4.0$ & $0.089 \pm 0.010$\\
E	& $13.23$ 		& $0.72$ & $0.32$ & $0.19$ & $0.44$ & $3.58 \pm 0.10$	& $2.5$ & $0.083 \pm 0.008$\\
F	& $13.15$ 		& $0.70$ & $0.29$ & $0.13$ & $0.44$ & $3.30 \pm 0.08$ & $5.0$ & $0.087 \pm 0.012$\\
\hline
\hline
\end{tabular}
\end{center}
\caption{Optimal parameters for each magneto-thermal evolution model (defined in Table~\ref{tab:par_rad}) that fit the observed $\log{N}$ - $\log{S}$ distribution.
The parameter $\eta$ (see eq.~\ref{eq:p_obs}) is also shown.}
\label{tab:par_X}
\end{table*}

Below $10^{-12}$ erg s$^{-1}$ cm$^{-2}$, the observed number of sources
saturates due to the sensitivity limit of the present instruments, combined with the interstellar medium absorption, which
only allows us to observe dim sources if they are nearby. 
Thus, we conclude that the degeneracy shown in Fig.~\ref{fig:b0}, where
a large region of the parameter space was allowed by the radio-pulsar data, is severely reduced 
when the thermally emitting X-ray pulsars are included in the analysis. Interestingly, for different models we obtain similar 
solutions for the initial magnetic field distribution, in the narrow region $\mu_{B_0}=13.15-13.35$,
$\sigma_{B_0} \approx 0.7-0.8$ and $\alpha=0.43-0.44$.

However, to reproduce the $\log{N}$-$\log{S}$ diagram is a necessary, but not sufficient condition to test a model.
We now turn to explore in more detail the particular form of the period distributions.
We remind again that the comparison is made with an incomplete observational data set, because the number of observed 
sources is only a fraction of the potentially detectable number of  NSs.  In Fig. \ref{fig:X} (bottom)  we show spin period 
histograms comparing the observed sample (black solid line) 
with synthetic models.

An important result can already be seen in these distributions. 
On one hand, one needs a significant fraction of magnetars/high-B field NSs in the initial distribution to account for the number of X-ray sources already present in our observed sample, 
but on the other hand, if that fraction is too large, the model over-predicts the number of X-ray pulsars with long periods, 
even assuming the existence of a highly resistive layer in the inner crust of NSs \citep{Pons2013}.
We can adjust the maximum period reached by a NS by varying the impurity parameter in the innermost part of the crust.
Increasing the value of $Q_{\rm imp}^{\rm ic}$ results in the decrease of the maximum period, 
but still not enough to successfully reproduce the observed upper limit of 12 s and, at the same time, explain the observed number
of very bright sources.
The strong constraint resulting from combining the need of a few very bright sources with 
the lack of isolated X-ray pulsars with periods longer than 12 s turns out to be a very restrictive condition.
The next relevant question is whether or not unknown observational selection effects are causing this discrepancy.

\begin{figure}
\begin{center}
\includegraphics[width=8.5cm]{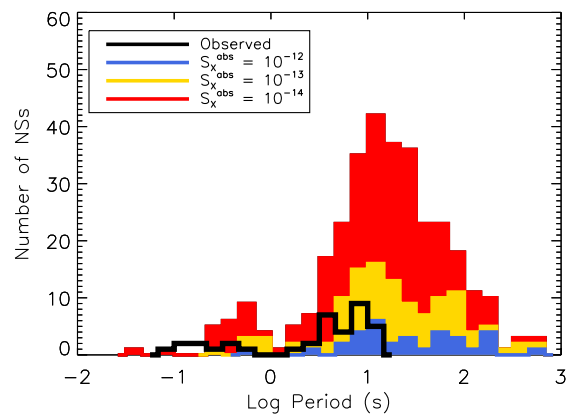}
\end{center}
\caption{Same as Fig.~\ref{fig:X} ({\it bottom}) for model D considering different flux thresholds:  $10^{-12}$ (blue), $10^{-13}$ (yellow) and $10^{-14}$ (red).
All fluxes are in units of erg s$^{-1}$ cm$^{-2}$.}
\label{fig:Flux}
\end{figure}

To explore how the period distribution varies with flux, let us focus on model D (the other models are qualitatively similar).
In Fig. \ref{fig:Flux} we compare the observed distribution to the synthetic data
with a different cut-off in the absorbed flux ($10^{-14}$, $10^{-13}$, and $10^{-12}$ erg s$^{-1}$ cm$^{-2}$).
An interesting trend is the apparent bimodal distribution of periods, similar to the observed distribution, 
although with a different scale and position. There is a partial lack of NSs with $P=1-2$ s, as seen in the observations, 
despite the initial period distribution being a single Gaussian.  However, even with the highest flux cut-off  
($10^{-12}$ erg s$^{-1}$ cm$^{-2}$) there are always too many visible X-ray pulsars with long periods. Thus, this discrepancy
cannot be attributed only to selection effects.

Since it is impossible to {\it a priori} take into account (theoretically) unknown selection effects in $X$-rays, 
we propose in the following a phenomenological approach. 
A close inspection of Fig.~\ref{fig:X} shows that the observational sample appears to be 
complete (given all the assumptions made in paper and within statistical fluctuations) for fluxes
$S_{X}^{abs} \gtrsim 3 \times 10^{-12}$ erg s$^{-1}$ cm$^{-2}$. 
We have also empirically found that 
the ratio of the number of observed sources to the number of potentially visible sources in the synthetic sample scales linearly 
with the flux in the range
$10^{-13}-10^{-12}$ erg s$^{-1}$ cm$^{-2}$. If we assume that the probability to detect
an X-ray pulsar with a given flux is simply
\begin{equation}\label{eq:p_obs}
p_{\rm obs} = \min \left\{ \eta S_{-11}, 1 \right\}
\end{equation}
where $S_{-11}$ is the absorbed flux in units of $10^{-11}$ erg s$^{-1}$ cm$^{-2}$, we can obtain for each model the value of $\eta$
that reproduces the $\log{N} - \log{S^{abs}_{X}}$ diagram. It turns out that  $\eta \sim 3-5$ for all models (eight column in Table
\ref{tab:par_X}). This naive approach is
useful to estimate the number of missed sources by uncontrolled selection effects.

\begin{figure}
\begin{center}
\includegraphics[width=8.5cm]{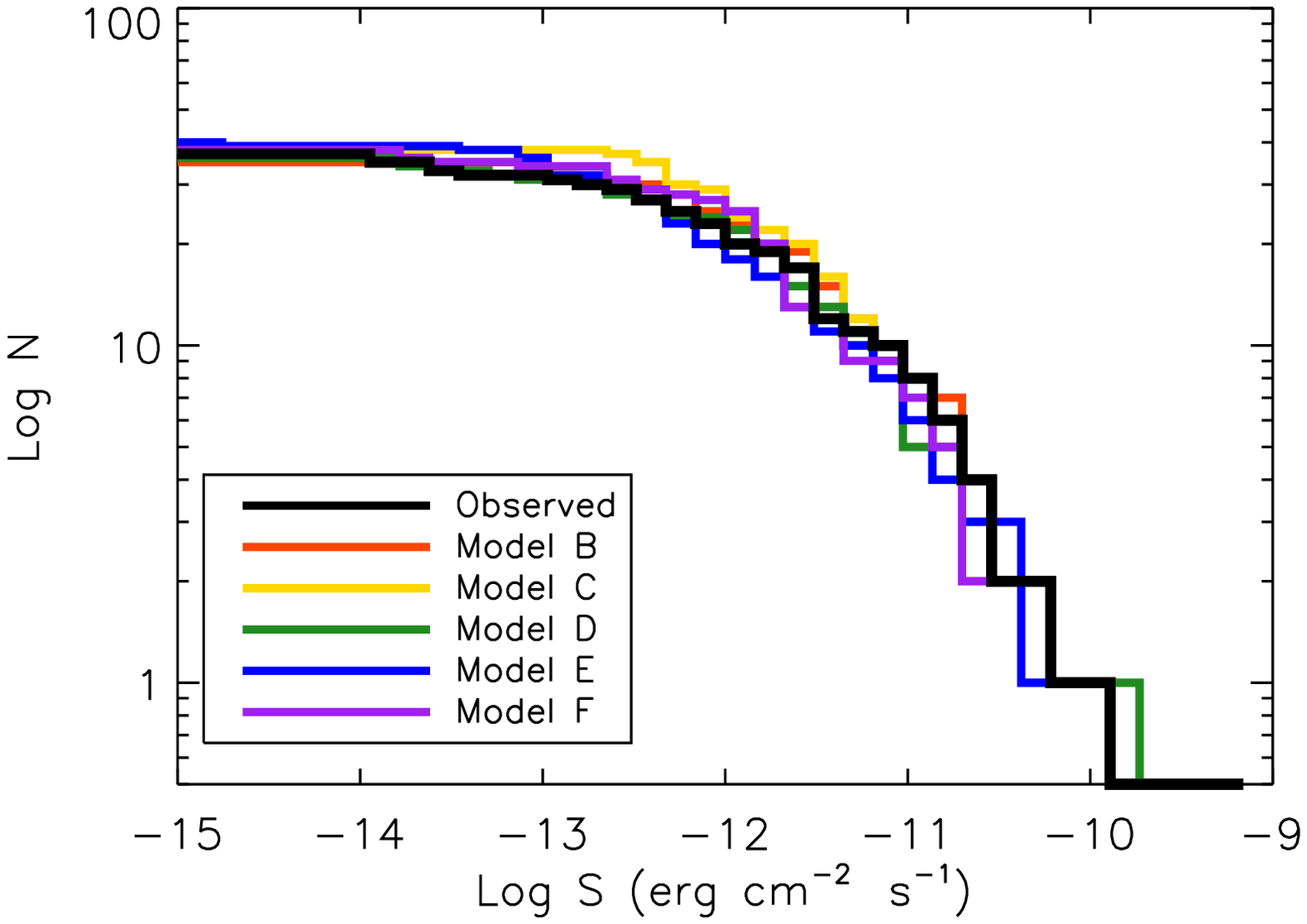}
\includegraphics[width=8.5cm]{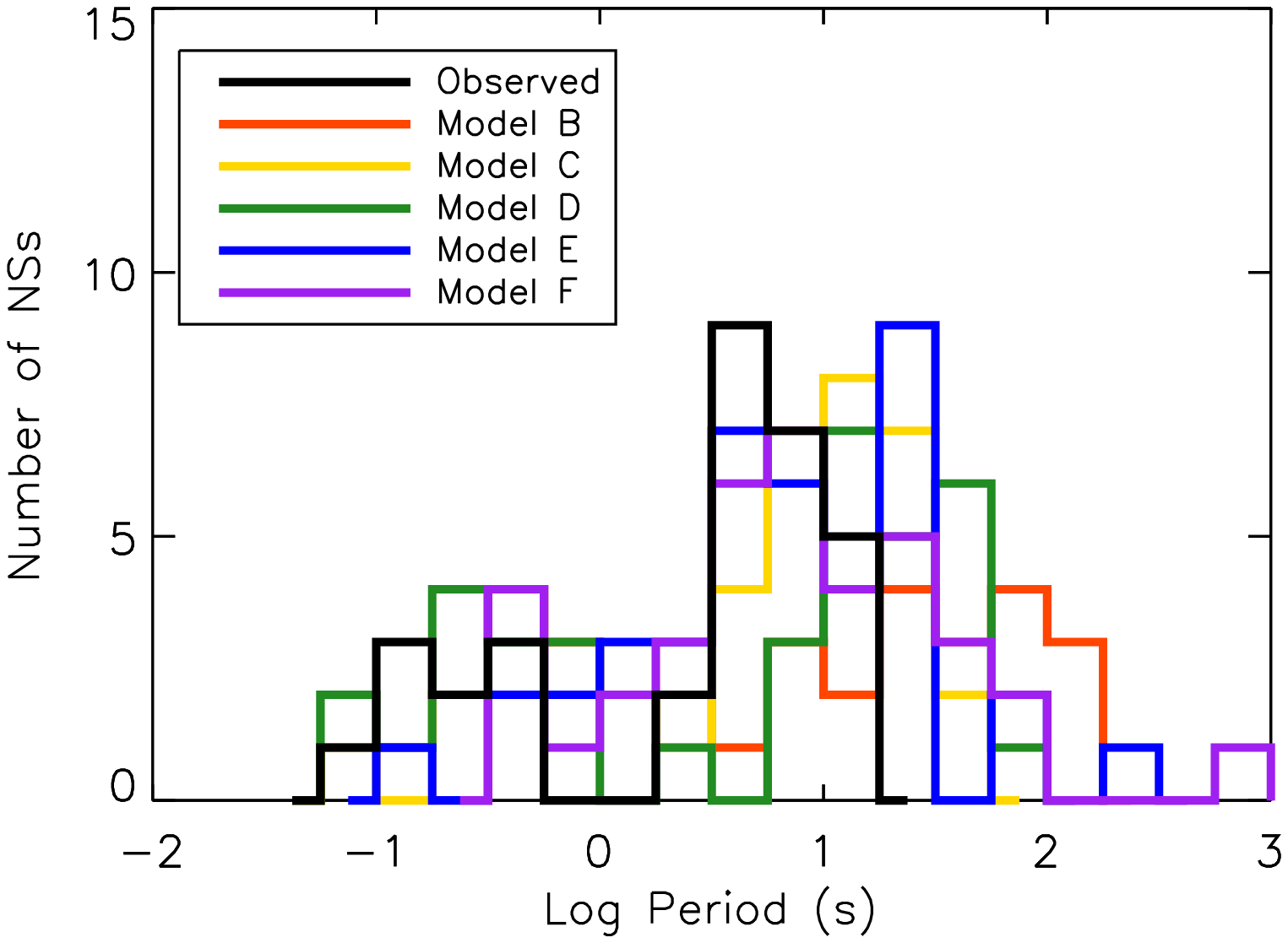}
\end{center}
\caption{Same as Fig.~\ref{fig:X}, but
applying the phenomenological detection probability  formula (eq.~\ref{eq:p_obs}).}
\label{fig:X_pobs}
\end{figure}

We can now generate a random {\it observed sample} from the synthetically generated sample by imposing that the 
probability to detect the thermal emission of a pulsar with a given flux is given by the above expression.
The corresponding $\log{N}$-$\log{S}$ diagrams and period distributions after such selection is applied 
are shown in  Fig.~\ref{fig:X_pobs}. 
Although most pulsars with high periods are now {\it not observed}, we still find a non-negligible number of bright 
enough sources with long periods. Our
predictions range from 4 to 20 X-ray pulsars to be observed with $P>12$ s, depending on the model. Even considering that 
we are dealing with low-number statistics, this appears to be a real problem.

\begin{table*}
\begin{center}
\begin{tabular}{|l|ccccccccr}
\hline
\hline
Model & $\mu_{B_0}$ 	& $\sigma_{B_0}$ & $\mu_{P_0}$  &  $\sigma_{P_0}$ & $\alpha$ & $n_{\rm br}$     &	$\eta$	&	${\cal{D}}$	\\
      & $\log{B}$ [G] 	& $\log{B}$ [G]  & [s]  	  &  [s]            &          & [century$^{-1}$] &		      &			\\
\hline
D	& $13.75$	& $1.15$ & $0.44$ & $0.26$ & $0.44$ & $6.81 \pm 0.20$ & $3.1$ & $0.120 \pm 0.012$ \\
E	& $13.58$	& $0.90$ & $0.33$ & $0.21$ & $0.44$ & $5.31 \pm 0.19$	& $5.0$ & $0.091 \pm 0.009$ \\
F	& $13.33$	& $0.83$ & $0.30$ & $0.19$ & $0.43$	& $3.24 \pm 0.09$	& $5.0$ & $0.110 \pm 0.008$ \\

\hline
\hline
\end{tabular}
\end{center}
\caption{Optimal parameters for models D, E and F with a truncated distribution of initial magnetic fields ($B_{0max} =  5 \times 10^{14}$G).}
\label{tab:par_X_b0cut}
\end{table*}

\subsection{Beyond Gaussian distributions of initial magnetic fields.}\label{sec:gauss}

\begin{figure}
\begin{center}
\includegraphics[width=8.5cm]{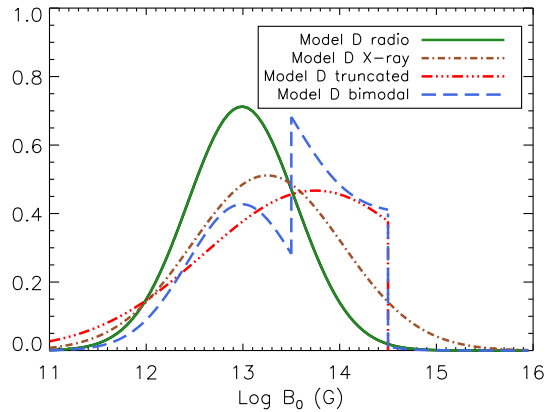}
\end{center}
\caption{$B_0$ distributions for model D 
that best fit the radio-pulsar (green/solid line) and X-ray pulsar (brown/dot-dashed) populations.
In red/three-dot dashed and blue/dashed we plot the results for truncated and bimodal distributions 
discussed in Section~\ref{sec:gauss}, respectively.}

\label{fig:b0hist}
\end{figure}

All previous results (and most studies in the literature) rely on the assumption that the initial magnetic field distribution is log-normal.
As the initial magnetic field increases, old NSs reach longer periods and therefore
the narrower beam size produces a strong selection of radio-pulsars.
Therefore, the low field part of the distribution is what we can practically constrain using the radio-pulsar population, while the
high field part of the distribution remains unconstrained, because the distribution of high field radio-pulsars is highly degenerate with
other parameters (beaming angle, radio-luminosity correlation, etc.). 
We have seen that the optimal parameters for radio-pulsars underpredicts the number of X-ray 
sources with high fluxes. We hence need to increase the width of the Gaussian to generate more bright NSs with high magnetic fields, 
but the problem is that the  Gaussian becomes too wide, and we run into an overprediction of X-ray pulsars with periods $P>12$s, which has not been observed.
The easiest, and perhaps more obvious, way out is to relax the assumption of an initial log-normal magnetic field
distribution. For example, if we consider model D, we can visualize  the $B_0$ distribution functions for the best fit to the radio-pulsar population (green/solid) and X-ray populations (brown/dot-dashed) in Fig.~\ref{fig:b0hist}. The excess of long period X-ray pulsars is originated by the high $B_0$
tail of the latter. We now discuss two possibilities, to reconcile the synthetic models with the data.

\begin{figure}
\begin{center}
\includegraphics[width=8.5cm]{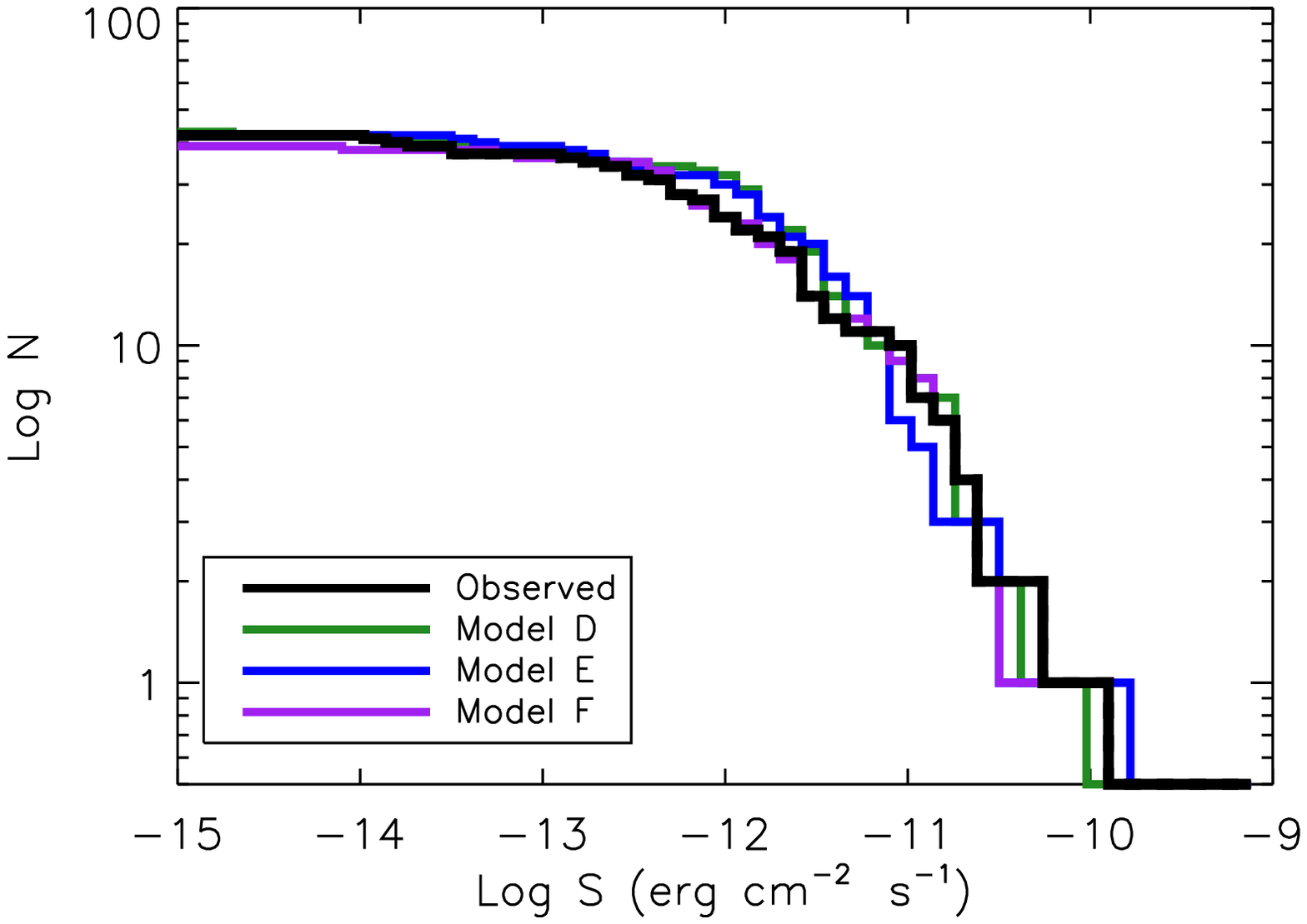}
\includegraphics[width=8.5cm]{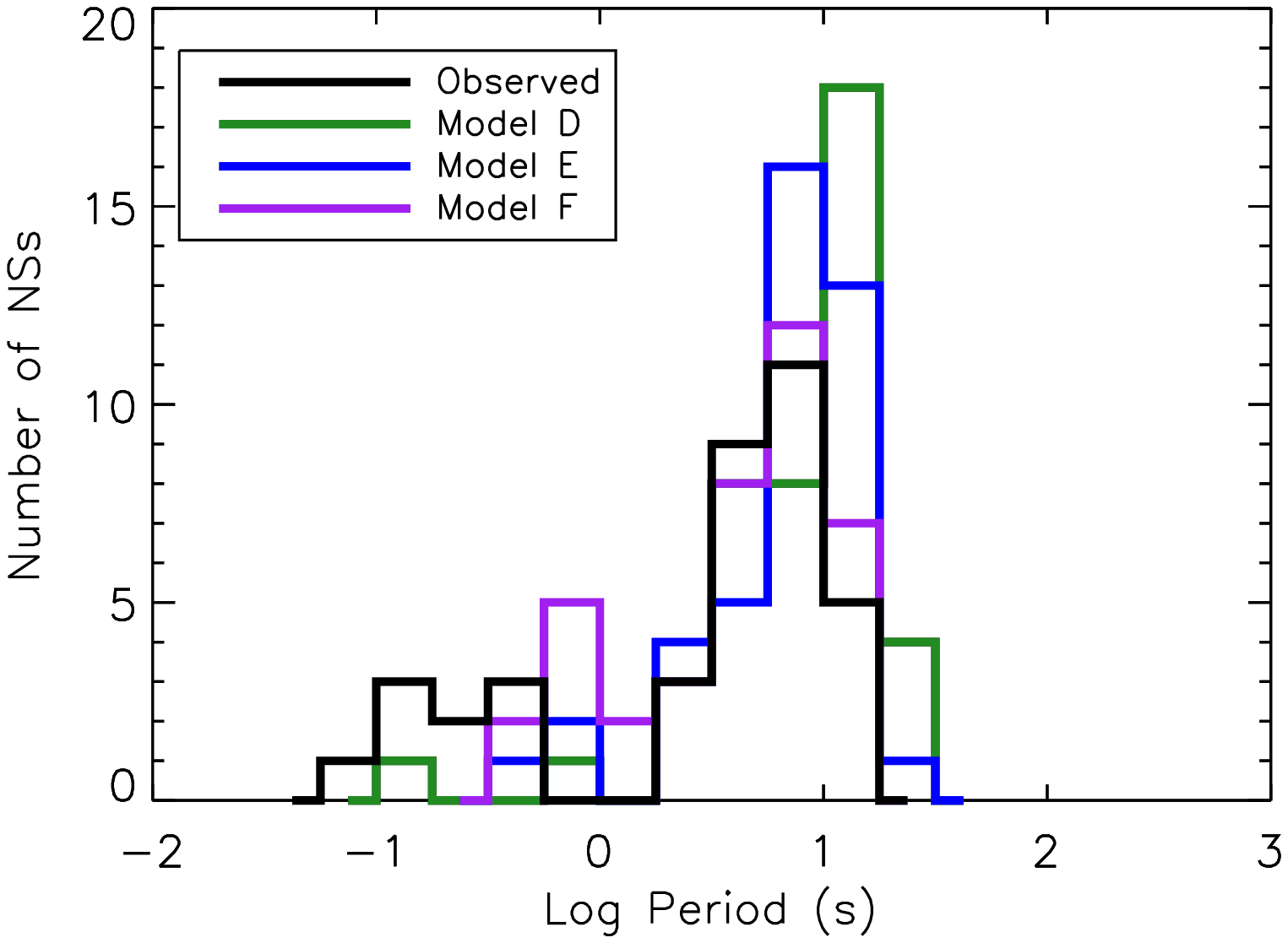}
\end{center}
\caption{Same as Fig.~\ref{fig:X_pobs} assuming a log-normal distribution with cutoff at $B_{0max} =  5 \times 10^{14}$ G,
and the parameters of Table~\ref{tab:par_X_b0cut}. 
}
\label{fig:X_b0cut}
\end{figure}

\subsubsection{A truncated magnetic field distribution.}\label{sec:truncated}

If the origin of the neutron star magnetic field is attributed to magneto-hydrodynamic instabilities (e.g., any dynamo process related to rotation or convection), a nonlinear saturation is expected when the amplified magnetic field becomes dynamically relevant to suppress the instability. 
This may happen at typical fields of the order of $10^{15}$ G, depending on the particular mechanism \citep{Ober2009,Ober2014}. 
Thus, we can simply introduce a cut-off value $B_{0max}$ to the log-normal distribution  (red/three-dot dashed curve in Fig.~\ref{fig:b0hist}). In Fig.~\ref{fig:X_b0cut}, 
we can see that the $\log{N}$-$\log{S}$ diagram for the X-ray population can also be successfully reproduced
by a truncated distribution with  $B_{0max} =  5 \times 10^{14}$G and increased $\mu_{B_0}$ and $\sigma_{B_0}$ 
(Table~\ref{tab:par_X_b0cut}).
The period distribution is significantly modified, and now there are no visible sources with $P>20$ s, as shown in Fig.~\ref{fig:X_b0cut}.

\begin{table}
\begin{center}
\begin{tabular}{|l|cccccr}
\hline
\hline
Model &	$NS1$	&	$NS2$	& $n_{\rm br}$     &	$\eta$	&	${\cal{D}}$	\\
      &	(\%)  &     (\%)  & [century$^{-1}$] &		      &			\\
\hline
D	& $60$	&	$40$	& $4.34 \pm 0.13$	& $6.3$  & $0.087 \pm 0.007$  \\
E	& $65$	&	$35$	& $4.80 \pm 0.10$	& $2.0$  & $0.082 \pm 0.006$ \\
F	& $70$	&	$30$	& $3.78 \pm 0.08$	& $3.2$  & $0.081 \pm 0.005$ \\

\hline
\hline
\end{tabular}
\end{center}
\caption{Results for models D, E and F with a bimodal distribution. 
We show the relative weight of the first population ($NS1$, same parameters as in Table \ref{tab:par_rad}) 
and second population of highly-magnetized stars ($NS2$, flat distribution in the range $\log B_0 \in [13.5,14.5]$ ).}
\end{table}

\subsubsection{A second population of magnetised NSs.}\label{sec:bimod}

\begin{figure}
\begin{center}
\includegraphics[width=8.5cm]{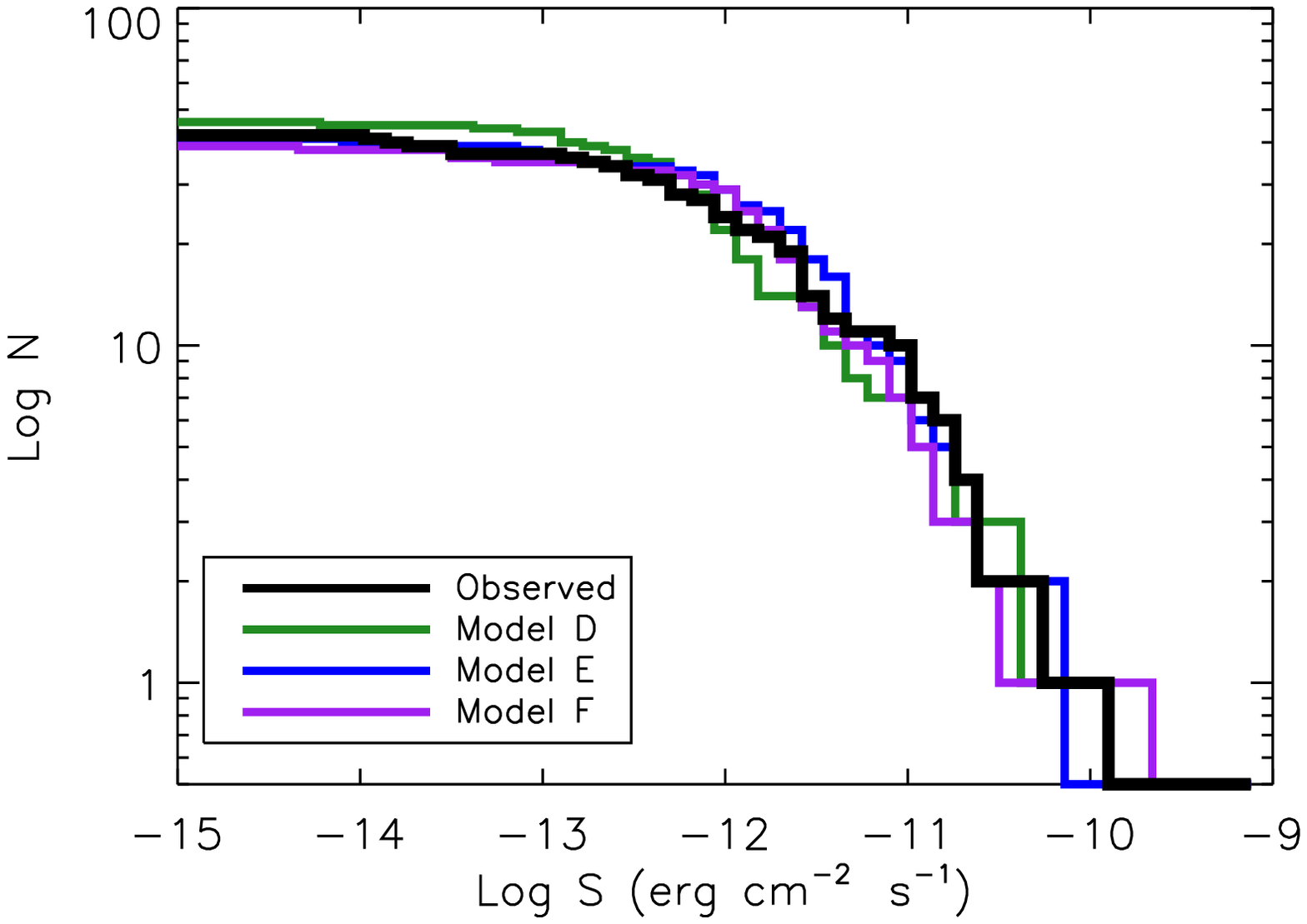}
\includegraphics[width=8.5cm]{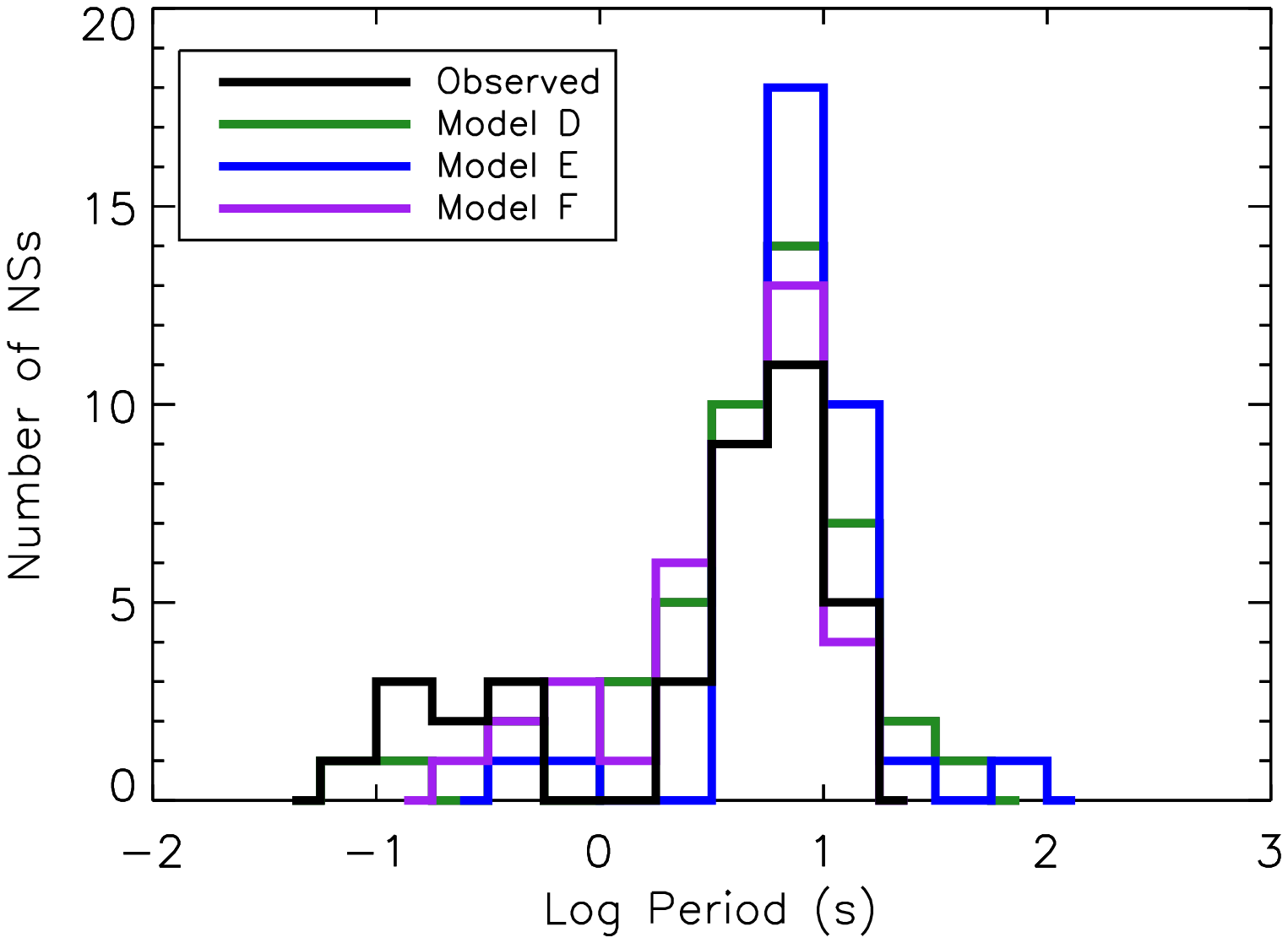}
\end{center}
\caption{Same as Fig.~\ref{fig:X_pobs}  with a bimodal distribution of $B_0$ described in \S 3.1.2.}
\label{fig:bimodal}
\end{figure}

The alternative possibility that we have explored is the existence of a second population of magnetised NSs, whose origin is 
different from  that of standard radio-pulsars. The reason for this second population could be attributed to 
two different evolutionary paths of  massive stars if they are isolated or in binary systems.
Since about 80\% of massive stars are in binaries, and assuming that half of them go through a common envelope phase 
or some evolutionary stage that produces a magnetized stellar core, one could expect about 40\% of the total pulsar population being
born with a binary massive progenitors, possibly leading to the formation of more magnetized NSs \citep{Spruit2008,Langer2012,Clark2014}.
We have investigated what happens if, in addition to the population generated with the parameters of Table~\ref{fig:lognlogs_rad} by fitting {\it only} the radio pulsar population (hereafter denoted by NS1),
we add a second population (denoted by NS2), uniformly distributed in the range $\log B_0 \in [13.5,14.5]$
and normalized to account for the total number of NSs observed (blue/dashed curve in Fig.~\ref{fig:b0hist}). 
The results are shown in Fig.~\ref{fig:bimodal}, where again we
can observe that both the $\log{N}$-$\log{S}$ diagram and the period distribution are satisfactorily reproduced.
We have verified that this second component does not modify the radio pulsar population, since their magnetic fields are 
so large that the vast majority of NS2 sources remains undetectable as radio-pulsars.
The total birthrate, however, will increase accordingly, still marginally compatible with the observations. 
We also note that one can obtain similar results  by assuming other distributions
for this second population, for example a narrow log-normal distribution centered  in $\approx 1-3 \times 10^{14}$G. 

\subsection{The upper limit to the birth rate of very strongly magnetized NSs.}

Before concluding our analysis on the period distribution, a simple calculation can stress once more the conflict
between having a significant fraction of NSs with magnetic fields at birth above a certain value, and the lack of observed
isolated NSs with periods $>12$ s. To quantify this incompatibility, we begin by obtaining the probability of observing a star, 
born with a given initial magnetic field $B_0$, as an X-ray pulsar with a period longer than 12 s. 
The results for three different values of the initial magnetic field are given in Table~\ref{tab:prob}, for the representative model D 
(similar values are obtained for other models). 
In this calculation, we assumed a uniform distribution of ages in the range 0-560 Myrs (with constant birth rate), 
and we generate a large number of NSs
with the same magnetic field. Then, we can obtain the probability as the fraction of synthetic {\it visible} sources with $P>12$ s 
to the total number of  stars generated. 

Next we address the following question: what is the maximum number of stars born with $B_0$
compatible (within statistical fluctuations) with the absence of observed neutron stars with periods longer than 12 s?  
This number can be estimated taking into account that if $N_{B_0}$ is the number of 
stars born with a given magnetic field, the probability of observing none of them with periods longer than 12 s is given by
$e^{-N_{B_0} p}$, according to the Poisson distribution, and the probability that we will observe {\it at least one} star with  $P>12$~s
is $1-e^{-N_{B_0} p}$. 

We can use this value to reject the null hypothesis in the following sense:
for a given model, we can calculate the value of $N_{B_0}$ that corresponds to a probability of $1-e^{-N_{B_0} p}=0.01$, 
and we know that in 99\% of the realizations with $N_{B_0}$ generated stars, we will not detect any source with $P>12$~s. 
Therefore, we can conclude that
if a model predicts $\ge N_{B_0}$ stars, the fact that none has been observed rejects the model at the 99\% confidence level
(statistical fluctuations may result in a no-detection only in $\le 1\%$ of the cases).

\begin{figure}
\begin{center}
\includegraphics[width=8.5cm]{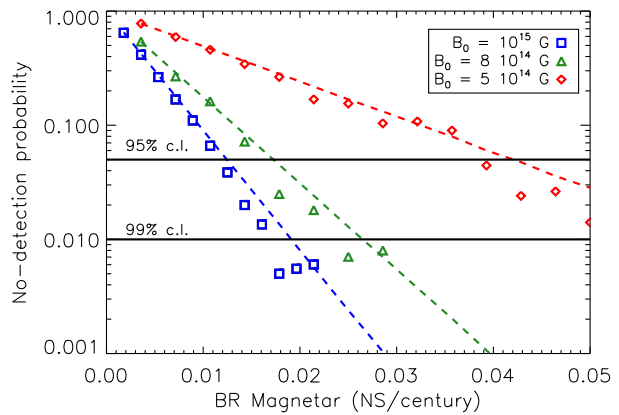}
\end{center}
\caption{Probability of no-detection of NSs with $P>12$ s as a function of the birthrate, for three  
different initial magnetic fields: $B_0 = 1 \times 10^{15}$ G (blue squares), $8 \times 10^{14}$ G (green triangles) 
and $5 \times 10^{14}$ G (red diamonds).
The symbols show results of simulations with 20 million NSs and the dashed lines correspond to the theoretical Poissonian value 
$e^{-N_{B_0} p}$.}
\label{fig:ndet}
\end{figure}

In Fig.~\ref{fig:ndet} we plot the probability of no-detection of NSs with periods longer than 12 s
against the birthrate, for different values of $B_0$. This can be interpreted as an upper limit to the birthrate of stars with initial 
magnetic field $>B_0$, for a given confidence level. 
For a confidence level of 99\% and a magnetic field of $10^{15}$ G we obtain
$N_{B_0}=112000$. This value corresponds to a birthrate of $\sim 0.02$ objects/century which means, by comparison 
with the total galactic NS birthrate $\sim 3$ NSs/century, that less than $0.7 \%$ can be born with magnetic fields 
 $\gtrsim10^{15}$ G. Of course, this fraction is even smaller if we reduce our confidence level.    
The maximum birthrate for stars with initial magnetic field greater than $5\times 10^{14}$ G can be $3$ times larger than for
$B_0=10^{15}$ G for the same confidence level.
This is satisfied by both alternative models described in the two subsections above, which cannot be discriminated with current data. 

\begin{table}
\begin{center}
\begin{tabular}{|l|c}
\hline
\hline
$B_0$ [G] & $p$ \\
\hline
$5 \times 10^{14}$ & $1.3 \times 10^{-5}$ \\
$8 \times 10^{14}$ & $3.1 \times 10^{-5}$ \\
$1 \times 10^{15}$ & $4.1 \times 10^{-5}$ \\

\hline
\hline
\end{tabular}
\end{center}
\caption{Probability of detection of stars whose period is longer than $12$ s for samples with fixed initial magnetic field $B_0$.}
\label{tab:prob}
\end{table}

\subsection{$\dot{P}$ distribution.}

\begin{figure}
\begin{center}
\includegraphics[width=8.5cm]{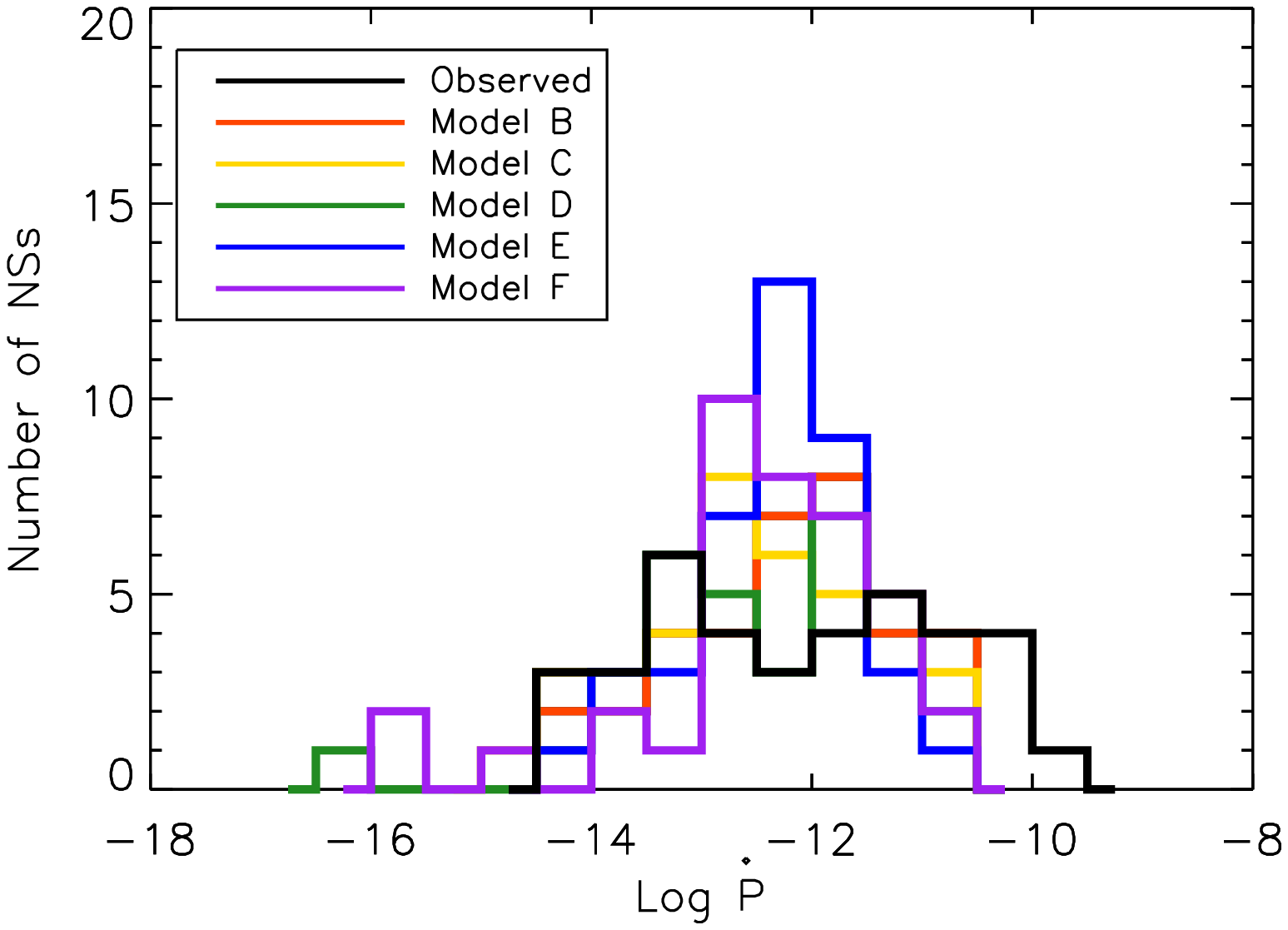}
\includegraphics[width=8.5cm]{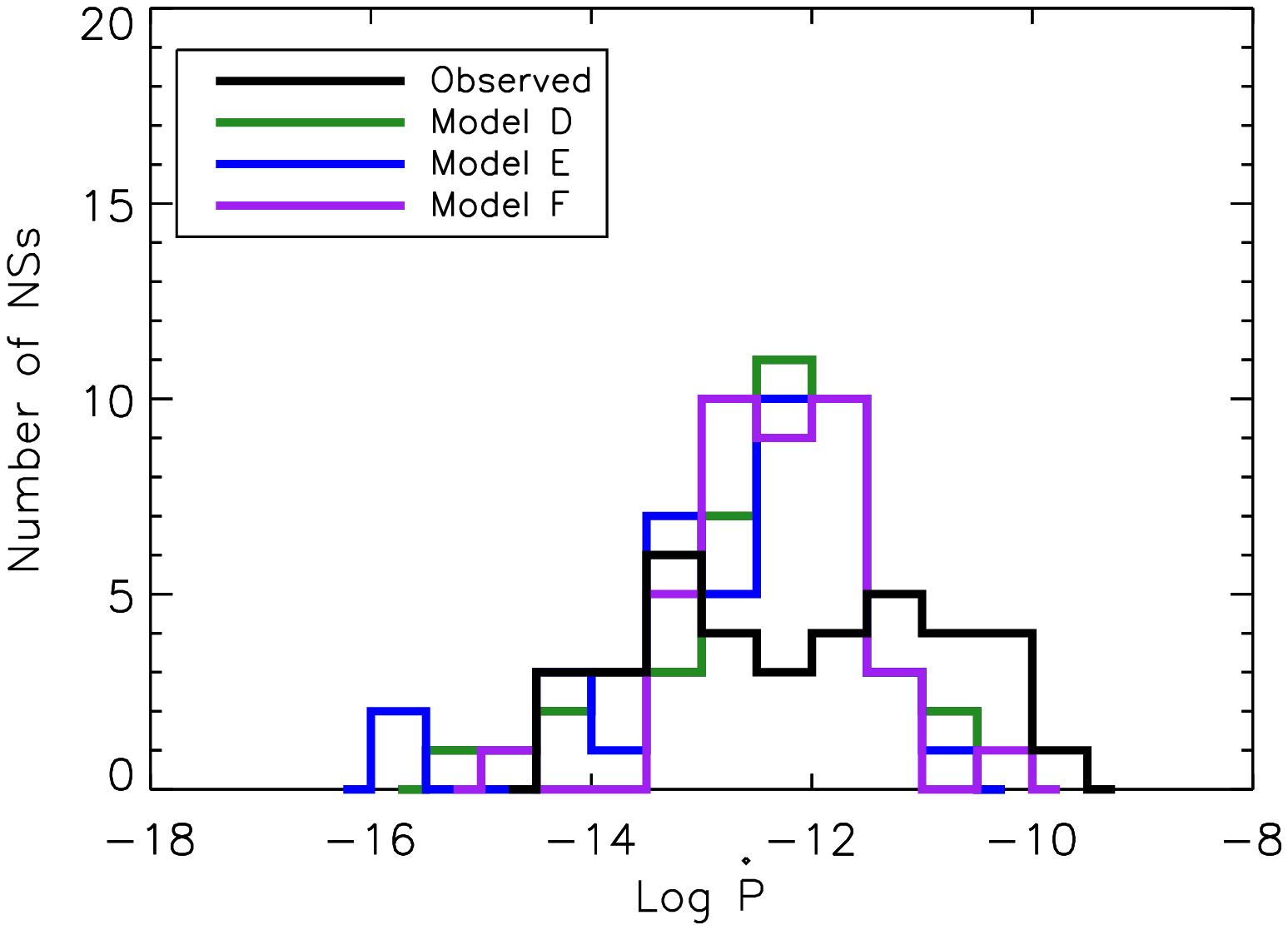}
\includegraphics[width=8.5cm]{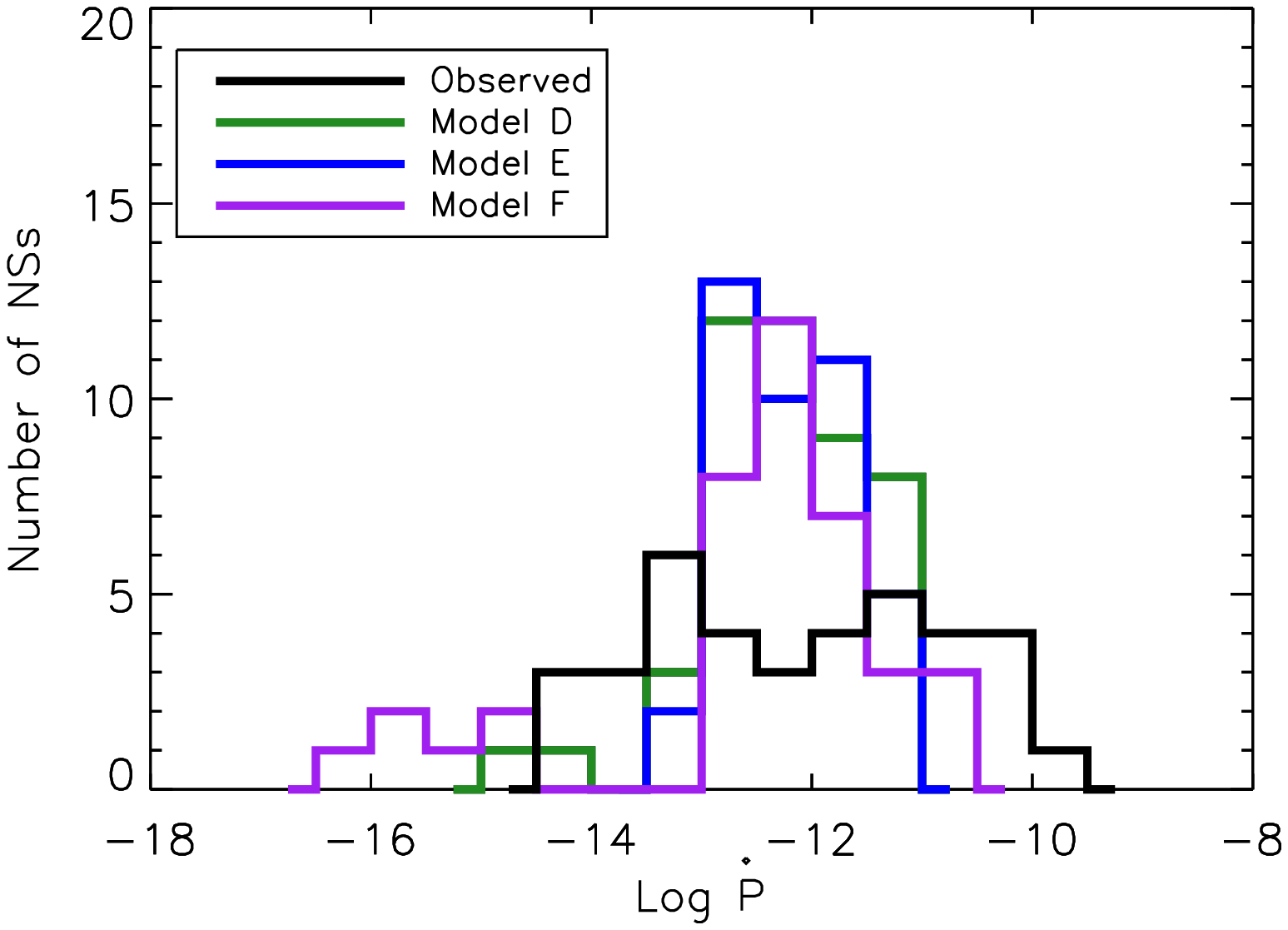}
\end{center}
\caption{Observed and predicted $\dot{P}$ distributions.
Top: models with log-normal distributions. 
Middle: models with truncated distributions discussed in section~\ref{sec:truncated}.
Bottom: models with bimodal distributions discussed in Section~\ref{sec:bimod}.
In colour lines we show the PS results while the black line shows the observed sources.}
\label{fig:pdhist_X_pobs}
\end{figure}

To complete our discussion, we comment on our results for the distribution of $\dot{P}$. 
In Fig.~\ref{fig:pdhist_X_pobs} we show the histograms for the models considered, which roughly agree with
the observations. However, all models and initial field distributions (log-normal, truncated log-normal, bimodal) show a similar trend: 
synthetic samples display an excess of sources with mean 
values $\sim 10^{-12}$ s s$^{-1}$ and a lack of a few sources with $\dot{P} > 10^{-11}$ s s$^{-1}$.
This is due to the fact that even the strongest magnetic fields ($B_0 \gtrsim 10^{15}$ G) hardly achieve $\dot{P} \sim 10^{-10}$ s s$^{-1}$.

The observed sources with very high $\dot{P}$ are magnetars and it is likely that modelling their spin evolution with the simple
magnetospheric torque formulae \citep{Spitkovsky,Beskin, Philippov} is an oversimplification.
For instance, strong particle winds can be non-negligible in the early $\sim 1000$ yr of their lives \citep{Windbraking} 
when these sources undergo frequent bursts and flares.
As a result of additional torques, the $\dot{P}$ can increase by an order of magnitude for the same magnetic field. This would imply
that the magnetic field estimate by the classical magnetodipolar formula is a factor of a few higher than the real value. This effect is only
expected to work during a short time and does not change the results of the analysis of the whole NS population, but it may contribute to 
the anomalous high values of  $\dot{P}$ of the few youngest magnetars.

\section{Summary}\label{sec:conclusions}

We have extended previous works on population synthesis of isolated neutron stars to study the
combination of initial conditions and microphysical parameters required to
explain the radio-pulsars and thermally emitting $X$-ray pulsar populations within the same evolutionary model.

Our first finding is that the  $\log{N}$-$\log{S}$ diagram of 
X-ray pulsars can be well fitted by several models that, at the same time, reproduce the radio-pulsar distribution
(because the parameter space of initial magnetic field distribution of radio-pulsars has a large region with the same 
statistical significance). Models with iron envelopes require a higher mean value of the initial magnetic field and a wider distribution 
than those with light elements. 

Our second relevant result is that the obtained sample of observed sources with 
$S_{X}^{abs} \gtrsim 3 \times 10^{-12}$ erg s$^{-1}$ cm$^{-2}$ is nearly complete, 
and we estimate that, with more sensitive instruments, the thermal emission of about one hundred of 
NSs with fluxes  $S^{abs}_{X} > 10^{-13}$ erg s$^{-1}$ cm$^{-2}$ is potentially detectable. 
This number can be increased up to 500  sources if the sensitivity is good enough to cover the whole sky with $ S^{abs}_{X}  > 10^{-14}$ 
erg s$^{-1}$ cm$^{-2}$.

However, when examining in detail the period distribution all models with log-normal distributions of the initial magnetic field fail to explain
why there are no X-ray pulsars with periods longer than 12 seconds. This sharp upper limit cannot be reproduced, even with 
models with a very high resistivity in the crust/core interface, if the initial field distribution is assumed to be log-normal. The problem
is solved if other distributions are invoked, such as a truncated log-normal distribution, or a binormal distribution with two
distinct populations. The most interesting feature, common to the two alternative cases considered, is that {\it both} cases require that 
magnetars cannot be born with a very high magnetic field.
For a confidence level of 99\%, we obtain an upper limit to the birth rate of NSs with  $B_0 > 10^{15}$ G of $\lesssim 0.02$ objects/century,
which represents a tiny fraction of the population (less than 1\%).

The existence of an upper limit to the magnetic field at birth in the range
$B_0 = 5-8 \times 10^{14}$ G, combined with fast magnetic field dissipation and observational selection effects, explains the lack of
isolated pulsars with long periods. Although this has a drawback, the lack of a few pulsars with very high values of 
$\dot{P}>10^{-11}$, note that we have assumed that NSs spindown only by the rotating 
magnetospheric torque.  In very young, active magnetars, losses of angular momentum carried by particles (wind) can compete
with the magnetospheric torque \citep{Windbraking}, increasing the value of $\dot{P}$ for a given magnetic field.
A more detailed study of these phenomena, and other modifications of the spin-down evolution is reserved to future works.

\section*{Acknowledgments}

This work was supported in part by the grants AYA2013-42184-P and Prometeu/2014/69, and by 
the {\it New Compstar} COST action MP1304. MG is supported by the fellowship BES-2011-049123.
DV and NR acknowledges support from grants AYA2012-39303 and SGR2014-1073. 
NR is additionally supported by an NWO Vidi award.  
We gratefully acknowledge useful discussions and suggestions from Peter Gonthier.

\bibliography{references}

\begin{thebibliography}{53}
\expandafter\ifx\csname natexlab\endcsname\relax\def\natexlab#1{#1}\fi

\bibitem[{{Arzoumanian}, {Chernoff} \& {Cordes}(2002){Arzoumanian}, {Chernoff},
  \& {Cordes}}]{ACC}
{Arzoumanian} Z., {Chernoff} D.~F., {Cordes} J.~M., 2002, \apj, 568, 289

\bibitem[{{Balucinska-Church} \& {McCammon}(1992)}]{phabs}
{Balucinska-Church} M., {McCammon} D., 1992, \apj, 400, 699

\bibitem[{{Beloborodov}(2013)}]{beloborodov13}
{Beloborodov} A.~M., 2013, \apj, 762, 13

\bibitem[{{Beskin}, {Istomin} \& {Philippov}(2013){Beskin}, {Istomin}, \&
  {Philippov}}]{Beskin}
{Beskin} V.~S., {Istomin} Y.~N., {Philippov} A.~A., 2013, Physics Uspekhi, 56,
  164

\bibitem[{{Cheng}, {Ho} \& {Ruderman}(1986){Cheng}, {Ho}, \&
  {Ruderman}}]{cheng86}
{Cheng} K.~S., {Ho} C., {Ruderman} M., 1986, \apj, 300, 500

\bibitem[{{Clark} {et~al}\mbox{.}(2014){Clark}, {Ritchie}, {Najarro}, {Langer},
  \& {Negueruela}}]{Clark2014}
{Clark} J.~S., {Ritchie} B.~W., {Najarro} F., {Langer} N., {Negueruela} I.,
  2014, \aap, 565, A90

\bibitem[{{Cordes} \& {Chernoff}(1998)}]{Cordes1998}
{Cordes} J.~M., {Chernoff} D.~F., 1998, \apj, 505, 315

\bibitem[{{Duncan} \& {Thompson}(1992)}]{duncan92}
{Duncan} R.~C., {Thompson} C., 1992, \apjl, 392, L9

\bibitem[{{Emmering} \& {Chevalier}(1989)}]{Emmering1989}
{Emmering} R.~T., {Chevalier} R.~A., 1989, \apj, 345, 931

\bibitem[{{Fasano} \& {Franceschini}(1987)}]{KS2D}
{Fasano} G., {Franceschini} A., 1987, \mnras, 225, 155

\bibitem[{{Faucher-Gigu{\`e}re} \& {Kaspi}(2006)}]{Faucher}
{Faucher-Gigu{\`e}re} C.-A., {Kaspi} V.~M., 2006, \apj, 643, 332

\bibitem[{{Ferrario} \& {Wickramasinghe}(2006)}]{Ferrario2006}
{Ferrario} L., {Wickramasinghe} D., 2006, \mnras, 367, 1323

\bibitem[{{Gonthier}, {Van Guilder} \& {Harding}(2004){Gonthier}, {Van
  Guilder}, \& {Harding}}]{Gon2004}
{Gonthier} P.~L., {Van Guilder} R., {Harding} A.~K., 2004, \apj, 604, 775

\bibitem[{{Gull{\'o}n} {et~al}\mbox{.}(2014){Gull{\'o}n}, {Miralles},
  {Vigan{\`o}}, \& {Pons}}]{Gullon2014}
{Gull{\'o}n} M., {Miralles} J.~A., {Vigan{\`o}} D., {Pons} J.~A., 2014, \mnras,
  443, 1891

\bibitem[{{Gunn} \& {Ostriker}(1970)}]{Gunn1970}
{Gunn} J.~E., {Ostriker} J.~P., 1970, \apj, 160, 979

\bibitem[{{Hartman} {et~al}\mbox{.}(1997){Hartman}, {Bhattacharya}, {Wijers},
  \& {Verbunt}}]{Hartman1997}
{Hartman} J.~W., {Bhattacharya} D., {Wijers} R., {Verbunt} F., 1997, \aap, 322,
  477

\bibitem[{{Kalapotharakos}, {Harding} \& {Kazanas}(2014){Kalapotharakos},
  {Harding}, \& {Kazanas}}]{Kalapo2014}
{Kalapotharakos} C., {Harding} A.~K., {Kazanas} D., 2014, \apj, 793, 97

\bibitem[{{Langer}(2012)}]{Langer2012}
{Langer} N., 2012, \araa, 50, 107

\bibitem[{{Lorimer} {et~al}\mbox{.}(1993){Lorimer}, {Bailes}, {Dewey}, \&
  {Harrison}}]{Lorimer1993}
{Lorimer} D.~R., {Bailes} M., {Dewey} R.~J., {Harrison} P.~A., 1993, \mnras,
  263, 403

\bibitem[{{Lyne}, {Manchester} \& {Taylor}(1985){Lyne}, {Manchester}, \&
  {Taylor}}]{Lyne1985}
{Lyne} A.~G., {Manchester} R.~N., {Taylor} J.~H., 1985, \mnras, 213, 613

\bibitem[{{Lyutikov} \& {Gavriil}(2006)}]{lyutikov06}
{Lyutikov} M., {Gavriil} F.~P., 2006, \mnras, 368, 690

\bibitem[{{Muslimov} \& {Harding}(2003)}]{muslimov03}
{Muslimov} A.~G., {Harding} A.~K., 2003, \apj, 588, 430

\bibitem[{{Narayan} \& {Ostriker}(1990)}]{Narayan1990}
{Narayan} R., {Ostriker} J.~P., 1990, \apj, 352, 222

\bibitem[{{Obergaulinger} {et~al}\mbox{.}(2009){Obergaulinger},
  {Cerd{\'a}-Dur{\'a}n}, {M{\"u}ller}, \& {Aloy}}]{Ober2009}
{Obergaulinger} M., {Cerd{\'a}-Dur{\'a}n} P., {M{\"u}ller} E., {Aloy} M.~A.,
  2009, \aap, 498, 241

\bibitem[{{Obergaulinger}, {Janka} \& {Aloy}(2014){Obergaulinger}, {Janka}, \&
  {Aloy}}]{Ober2014}
{Obergaulinger} M., {Janka} H.-T., {Aloy} M.~A., 2014, \mnras, 445, 3169

\bibitem[{{Olausen} \& {Kaspi}(2014)}]{olausen14}
{Olausen} S.~A., {Kaspi} V.~M., 2014, \apjs, 212, 6

\bibitem[{{Page} {et~al}\mbox{.}(2004){Page}, {Lattimer}, {Prakash}, \&
  {Steiner}}]{Page2004}
{Page} D., {Lattimer} J.~M., {Prakash} M., {Steiner} A.~W., 2004, \apjs, 155,
  623

\bibitem[{{Page} {et~al}\mbox{.}(2011){Page}, {Prakash}, {Lattimer}, \&
  {Steiner}}]{Page2011}
{Page} D., {Prakash} M., {Lattimer} J.~M., {Steiner} A.~W., 2011, Physical
  Review Letters, 106, 081101

\bibitem[{{Perna} \& {Pons}(2011)}]{PernaPons}
{Perna} R., {Pons} J.~A., 2011, \apjl, 727, L51

\bibitem[{{P{\'e}tri}(2012)}]{petri12}
{P{\'e}tri} J., 2012, \mnras, 424, 2023

\bibitem[{{Philippov}, {Tchekhovskoy} \& {Li}(2013){Philippov}, {Tchekhovskoy},
  \& {Li}}]{Philippov}
{Philippov} A., {Tchekhovskoy} A., {Li} J.~G., 2013, arXiv:astro-ph/1311.1513

\bibitem[{{Phinney} \& {Blandford}(1981)}]{Phinney1981}
{Phinney} E.~S., {Blandford} R.~D., 1981, \mnras, 194, 137

\bibitem[{{Pierbattista} {et~al}\mbox{.}(2012){Pierbattista}, {Grenier},
  {Harding}, \& {Gonthier}}]{Pierbattista}
{Pierbattista} M., {Grenier} I.~A., {Harding} A.~K., {Gonthier} P.~L., 2012,
  \aap, 545, A42

\bibitem[{{Pons} \& {Perna}(2011)}]{PonsPerna}
{Pons} J.~A., {Perna} R., 2011, \apj, 741, 123

\bibitem[{{Pons}, {Vigan{\`o}} \& {Rea}(2013){Pons}, {Vigan{\`o}}, \&
  {Rea}}]{Pons2013}
{Pons} J.~A., {Vigan{\`o}} D., {Rea} N., 2013, Nature Physics, 9, 431

\bibitem[{{Popov} {et~al}\mbox{.}(2010){Popov}, {Pons}, {Miralles}, {Boldin},
  \& {Posselt}}]{PSB_Popov}
{Popov} S.~B., {Pons} J.~A., {Miralles} J.~A., {Boldin} P.~A., {Posselt} B.,
  2010, \mnras, 401, 2675

\bibitem[{{Potekhin}, {Chabrier} \& {Yakovlev}(2007){Potekhin}, {Chabrier}, \&
  {Yakovlev}}]{Potekhin2007}
{Potekhin} A.~Y., {Chabrier} G., {Yakovlev} D.~G., 2007, \apss, 308, 353

\bibitem[{Press {et~al}\mbox{.}(1993)Press, Teukolsky, Vetterling, \&
  Flannery}]{numericalrecipes}
Press W.~H., Teukolsky S.~A., Vetterling W.~T., Flannery B.~P., 1993, Numerical
  Recipes in FORTRAN; The Art of Scientific Computing, 2nd edn. Cambridge
  University Press, New York, NY, USA

\bibitem[{{Rea} {et~al}\mbox{.}(2008){Rea}, {Zane}, {Turolla}, {Lyutikov}, \&
  {G{\"o}tz}}]{rea08}
{Rea} N., {Zane} S., {Turolla} R., {Lyutikov} M., {G{\"o}tz} D., 2008, \apj,
  686, 1245

\bibitem[{{Romani}(1996)}]{romani96}
{Romani} R.~W., 1996, \apj, 470, 469

\bibitem[{{Ruderman} \& {Sutherland}(1975)}]{ruderman75}
{Ruderman} M.~A., {Sutherland} P.~G., 1975, \apj, 196, 51

\bibitem[{{Spitkovsky}(2006)}]{Spitkovsky}
{Spitkovsky} A., 2006, \apj, 648, L51

\bibitem[{{Spruit}(2008)}]{Spruit2008}
{Spruit} H.~C., 2008, in American Institute of Physics Conference Series, Vol.
  983, 40 Years of Pulsars: Millisecond Pulsars, Magnetars and More, {Bassa}
  C., {Wang} Z., {Cumming} A., {Kaspi} V.~M., eds., pp. 391--398

\bibitem[{{Stollman}(1987)}]{Stollman1987}
{Stollman} G.~M., 1987, \aap, 178, 143

\bibitem[{{Sturrock}(1971)}]{sturrock71}
{Sturrock} P.~A., 1971, \apj, 164, 529

\bibitem[{{Thompson} \& {Duncan}(1993)}]{thompson93}
{Thompson} C., {Duncan} R.~C., 1993, \apj, 408, 194

\bibitem[{{Thompson}, {Lyutikov} \& {Kulkarni}(2002){Thompson}, {Lyutikov}, \&
  {Kulkarni}}]{thompson02}
{Thompson} C., {Lyutikov} M., {Kulkarni} S.~R., 2002, \apj, 574, 332

\bibitem[{{Tong} {et~al}\mbox{.}(2013){Tong}, {Xu}, {Song}, \&
  {Qiao}}]{Windbraking}
{Tong} H., {Xu} R.~X., {Song} L.~M., {Qiao} G.~J., 2013, \apj, 768, 144

\bibitem[{{Vigan{\`o}} {et~al}\mbox{.}(2013){Vigan{\`o}}, {Rea}, {Pons},
  {Perna}, {Aguilera}, \& {Miralles}}]{Vigano2013}
{Vigan{\`o}} D., {Rea} N., {Pons} J.~A., {Perna} R., {Aguilera} D.~N.,
  {Miralles} J.~A., 2013, \mnras, 434, 123

\bibitem[{{Vranesevic} {et~al}\mbox{.}(2004){Vranesevic}, {Manchester},
  {Lorimer}, {Hobbs}, {Lyne}, {Kramer}, {Camilo}, {Stairs}, {Kaspi}, {D'Amico},
  {Possenti}, {Crawford}, {Faulkner}, \& {McLaughlin}}]{Vranesevic2004}
{Vranesevic} N. {et~al.}, 2004, \apjl, 617, L139

\bibitem[{{Yakovlev} \& {Pethick}(2004)}]{Yakovlev2004}
{Yakovlev} D.~G., {Pethick} C.~J., 2004, \araa, 42, 169

\bibitem[{{Zane} {et~al}\mbox{.}(2009){Zane}, {Rea}, {Turolla}, \&
  {Nobili}}]{zane09}
{Zane} S., {Rea} N., {Turolla} R., {Nobili} L., 2009, \mnras, 398, 1403

\bibitem[{{Zane} {et~al}\mbox{.}(1995){Zane}, {Turolla}, {Zampieri}, {Colpi},
  \& {Treves}}]{nh}
{Zane} S., {Turolla} R., {Zampieri} L., {Colpi} M., {Treves} A., 1995, \apj,
  451, 739

\end{thebibliography}

\label{lastpage}

\end{document}